\documentclass[aps,superscriptaddress,preprintnumbers,12pt,nofootinbib]{revtex4} 
\usepackage{graphicx}  
\usepackage{dcolumn}   
\usepackage{bm}        
\usepackage{amsmath}   
\usepackage{amsfonts}  
\usepackage{subfigure} 
\usepackage{hyperref}
\usepackage{color}
\usepackage{slashed}    
\newcommand{\be}{\begin{equation}}
\newcommand{\ee}{\end{equation}}
\def\ltap{\ \raise.3ex\hbox{$<$\kern-.75em\lower1ex\hbox{$\sim$}}\ }
\def\gtap{\ \raise.3ex\hbox{$>$\kern-.75em\lower1ex\hbox{$\sim$}}\ }
\def\lsim{\ \raise.3ex\hbox{$<$\kern-.75em\lower1ex\hbox{$\sim$}}\ }
\def\gsim{\ \raise.3ex\hbox{$>$\kern-.75em\lower1ex\hbox{$\sim$}}\ }
\def\eg{{\it e.g.}}
\def\ie{{\it i.e.}}
\newcommand{\met}{\slashed {E}_{T}}
\usepackage{pstricks}

\begin{document}
\preprint{FERMILAB-PUB-12-066-T}

\title{Taking a Razor to Dark Matter Parameter Space at the LHC}

\author{Patrick J. Fox}
\email{pjfox@fnal.gov}
\affiliation{Theoretical Physics Department, Fermilab, P.O. Box 500, Batavia, IL 60510, USA}

\author{Roni Harnik}
\email{roni@fnal.gov}
\affiliation{Theoretical Physics Department, Fermilab, P.O. Box 500, Batavia, IL 60510, USA}

\author{Reinard Primulando}
\email{rprimulando@email.wm.edu}
\affiliation{Theoretical Physics Department, Fermilab, P.O. Box 500, Batavia, IL 60510, USA}
\affiliation{High Energy Theory Group, Department of Physics, College of William and Mary, Williamsburg, VA 23187, USA}

\author{Chiu-Tien Yu}
\email{cyu27@wisc.edu}
\affiliation{Theoretical Physics Department, Fermilab, P.O. Box 500, Batavia, IL 60510, USA}
\affiliation{Department of Physics, University of Wisconsin, Madison, WI 53706 USA}

\date{\today}
\begin{abstract}
Dark matter (DM) has been searched for at colliders in a largely model independent fashion by looking for an excess number of events involving a single jet, or photon, and missing energy.  We investigate the possibility of looking for excesses in more inclusive jet channels.  Events with multiple jets contain more information and thus more handles to increase the signal to background ratio. 
In particular, we adapt the recent CMS ``razor'' analysis 
from a search for supersymmetry
to a search for DM and estimate the potential reach.  The region of razor variables which are most sensitive to dark matter are not covered by the current SUSY search.
We consider simplified models where DM is a Dirac fermion that couples to the quarks of the Standard Model (SM) through exchange of vector or axial-vector mediators or to gluons through scalar exchange.  We consider both light and heavy (leading to effective contact interactions) mediators.  Since the razor analysis requires multiple jets in the final state, the data set is complementary to that used for the monojet search and thus the bounds can be combined.
\end{abstract}
\maketitle
\newpage
\section{Introduction}\label{sec:intro}

Through precision cosmological measurements, we have uncovered many of the general properties of dark matter (DM) in the cosmos.  However, further determinations of the properties of DM and its distribution throughout the universe will require probing beyond its gravitational interactions.  Although there is considerable effort underway to indirectly observe DM through the signatures of DM annihilations in places of high expected density, such as the centers of our galaxy, galaxy clusters and dwarf galaxies, there is no substitute for detection of DM in a controlled lab setting.  To this end, there are many experiments presently searching for direct observation of DM scattering off nuclei in underground labs.  Intriguingly, both indirect and direct searches are finding interesting anomalies that are consistent with what is expected from DM.  Unfortunately, there is also considerable confusion since many of these excesses could also be consistent with backgrounds or systematic effects.  Furthermore, both the indirect and direct search techniques rely on inputs from astrophysics, such as the spatial and velocity distribution of the DM in our galaxy, or the spectrum and morphology of high energy gamma and cosmic rays, which are notoriously difficult to estimate.  

High energy colliders provide an alternative~\cite{Birkedal:2004xn}, complementary way to search for DM that is independent of assumptions about astrophysical quantities.  If DM is to be found in direct detection experiments then it must couple to quarks or gluons, and thus it is possible to directly produce DM in high energy hadron colliders.  Since DM carries no SM charge, it will leave the detector without further interactions, resulting in a missing (transverse) energy signature ($\met$).  Thus, the observation of an excess of events in channels involving missing energy could provide tantalizing evidence of the production of DM, and from these channels, DM properties such as its mass could be determined.  Similarly, if there are no observed excesses, one can place limits on the size of putative DM-quark/gluon couplings.  These collider limits can be re-expressed as a limit on DM-nucleon couplings and compared to the limits that come from the absence of events in dedicated direct detection experiments such as CDMS~\cite{Ahmed:2009zw} and XENON100~\cite{Aprile:2011hi}.

Many models of beyond the standard model (BSM) physics contain a viable DM candidate, and thus predict events involving $\met$.  Many ingenious search strategies have been developed within the context of particular models, but these strategies often rely on other unique and unrelated features specific to the model.  Furthermore, without independent evidence for any of these models, and armed only with the knowledge that DM exists, it is worthwhile to consider more model independent search strategies.  The simplest final state that could involve the production of DM and serve as a limit on its couplings is a monojet/monophoton in association with missing energy.  At the Tevatron, a search for $j+\met$ that was originally designed to search for large extra dimensions~\cite{CDFmonojet,Aaltonen:2008hh} has been recast as a constraint on DM production, both through contact interactions of DM and the SM~\cite{Goodman:2010yf,Bai:2010hh,Goodman:2010ku}, and through the presence of a light mediator particle~\cite{Bai:2010hh,Goodman:2011jq,Shoemaker:2011vi}.  These analyses were based on $\sim 1$ fb$^{-1}$ of data and a simple cut-and-count approach.  Recently, CDF has carried out a dedicated search for DM in the monojet channel, using 6.7 fb$^{-1}$ and the full shape information contained in the monojet spectrum~\cite{Aaltonen:2012jb}.  For heavy DM, these bounds can be improved upon by going to the LHC, and analyses of monojets~\cite{Rajaraman:2011wf,Fox:2011pm,Shoemaker:2011vi} and monophotons~\cite{Fox:2011pm} have been carried out on $\sim 1$ fb$^{-1}$ of data.  Very recently CMS has released a DM search in the monophoton channel~\cite{CMSmonophoton}.
Constraints from LEP monophoton and missing energy searches have also been calculated~\cite{Fox:2011fx,Fortin:2011hv}.

Although the monojet/monophoton is certainly the simplest final state one can expect to find DM, it does not necessarily result in the strongest limits\footnote{As has recently been discussed~\cite{An:2012va}, if there is a light mediator coupling the SM to DM, searches for the mediator in the dijet channel are a complementary way to constrain the DM and its couplings.}.  At the high collision energies typical of the LHC, one expects a hard process to be accompanied by several high $p_T$ jets, and the veto required to fit into the one jet topology may restrict the signal efficiency.  In addition, events with multiple jets contain more information, such as inter-jet angles. As we shall see, optimizing searches with respect to these variables may improve the ratio of signal to background efficiencies.
There are approaches such as the CMS ``monojet'' search~\cite{CMS-PAS-EXO-11-059} which allow a second hard jet as long as the topology is sufficiently far from back-to-back that QCD backgrounds are suppressed.  We take this philosophy one step further and investigate a more inclusive search approach that allows an arbitrary number of hard jets, as long as there is also considerable missing energy, see also~\cite{Beltran:2010ww}.  We base our strategy around that used by the CMS ``razor'' analysis~\cite{Rogan:2010kb,Chatrchyan:2011ek}, which was originally employed to search for supersymmetry, and was based on approximately 800 pb$^{-1}$ of data~\cite{CMSrazor}. 

This paper is outlined as follows.  In Sec.~\ref{sec:setup}, we introduce both the effective theory of DM coupling to quarks through contact operators, and some simplified models which UV complete these by introducing a mediator light enough to be accessible at the LHC. We describe the razor analysis in Sec.~\ref{sec:razor}, beginning with a description of the analysis in Sec.~\ref{sec:variables}.  In Sec.~\ref{sec:results}, we outline our results for the case of contact operators and in Sec.~\ref{sec:directdetection}, we compare the collider bounds with direct detection bounds. Finally, we address the issues that arise with light mediators and the validity of using an effective theory in~Sec.~\ref{sec:lightmediators}. 

\section{A Simplified Model of Dark Matter Interactions}\label{sec:setup}

As mentioned above, searches for DM in many models of BSM physics utilize additional features of the model, such as production of colored states that ultimately decay to DM.  Here, we wish to follow an approach that is more model independent and we introduce simplified models~\cite{Alves:2011wf} that couple DM to the SM.   In addition to the SM, these models contain the DM, $\chi$, which we assume to be a Dirac fermion~\footnote{This choice has little effect on our results, although the vector coupling would not be allowed for the case of Majorana DM.}, and a mediator particle that couples to the DM and states in the SM.  The nature of the mediator will determine the form of the SM-DM coupling and whether the non-relativistic limit is spin-independent (SI) or spin-dependent (SD).  We will consider vector, axial-vector, and scalar mediators, which give a representative sample of the different behaviors possible at colliders and direct detection experiments; for a more complete list of possibilities see for example~\cite{Goodman:2010ku,Cheung:2012gi}\footnote{We do not consider the scalar operator, $(\bar q q) (\bar \chi \chi)$, since this type of operator is suppressed by a Yukawa coupling. As a result the limits on $\Lambda$ are expected to be weak and in a region where the effective theory is not valid~\cite{Fox:2011pm}.}. The interaction Lagrangians for these mediators are given by:
\begin{eqnarray}
\mathcal{L}_V&=&g_\chi \, \bar\chi\gamma_\mu\chi \, Z'^\mu + g_q \, \bar q\gamma_\mu q \, Z'^\mu\, , \label{eq:LV}\\
\mathcal{L}_A&=&g_\chi \, \bar\chi\gamma_\mu\gamma_5\chi \, Z'^\mu + g_q \, \bar q\gamma_\mu\gamma_5 q \, Z'^\mu, ,\\
\mathcal{L}_G&=&g_\chi \, \bar\chi\chi \, S + \alpha_s \frac{S}{F}G^a_{\mu\nu}G^{a\mu\nu} \, , \label{eq:LG}
\end{eqnarray}
where $q$ is a SM quark field, $G_{\mu\nu}^a$ is the gluon field strength tensor, $Z'$ denotes a spin-1 mediator and $S$ denotes a spin-0 mediator.

We start by considering the limit of the simplified model where only the DM is accessible at colliders~\cite{Beltran:2010ww}, and the mediator is integrated out.  In this limit, with very heavy mediators ($\gtrsim$ few TeV), we can use the framework of effective field theory.  The resulting effective operators for each choice of mediator are:
\begin{eqnarray}
\mathcal{O}_V&=&\frac{(\bar\chi\gamma_\mu\chi)(\bar q\gamma^\mu q)}{\Lambda^2}\, , \label{eq:OV}\\
\mathcal{O}_A&=&\frac{(\bar\chi\gamma_\mu\gamma_5\chi)(\bar q\gamma^\mu\gamma_5 q)}{\Lambda^2}\, ,\\
\mathcal{O}_G&=&\alpha_s\frac{(\bar\chi\chi)(G^a_{\mu\nu}G^{a\mu\nu})}{\Lambda^3}\, , \label{eq:OG}
\end{eqnarray}
where $\Lambda^2 = M_{Z'}^2 / g_\chi g_q $ in both $\mathcal{O}_V$ and $\mathcal{O}_A$, and $\Lambda^3 = F M_S^2 / g_\chi$ for  $\mathcal{O}_G$. In Sec.~\ref{sec:lightmediators}, we will discuss whether this effective theory approach is valid and the effects of keeping the mediator in the simplified model.  We calculate the bounds for the up and down quarks separately, but the bound for any linear combination of quark flavors can be derived from these bounds~\cite{Fox:2011pm}. 
%

We ultimately want to compare collider bounds to direct detection bounds. Here, the effective theory in equations~(\ref{eq:OV})-(\ref{eq:OG}) is always valid.
In order to match the quark-level operators to nucleon-level operators, the coupling between the SM and DM must be of the form ${\cal{O}}_{SM}{\cal{O}}_\chi$, where ${\cal{O}}_{SM}$ contains only SM fields and ${\cal{O}}_\chi$ involves only DM such that we can extract the matrix element $\langle N|{\cal{O}}_{SM}|N\rangle$ \cite{Fan:2010gt}. At colliders, for a Dirac fermion $\chi$, both ${\cal{O}}_V$ and ${\cal{O}}_A$ contribute to $\chi$ production with roughly equal rates. However, in direct detection experiments, the spin-independent ${\cal{O}}_V$ dominates over the spin-dependent ${\cal{O}}_A$. ${\cal{O}}_V$ vanishes if we change our assumption to Majorana DM.

\section{Razor}\label{sec:razor}  

In this section, we estimate bounds on DM operators with the razor analysis. We begin with a description of the general razor analysis as used by CMS~\cite{CMSrazor}. We then compare the shape of signal and background events in the razor variables, $M_R$ and $R^2$, and identify cuts which are optimal for searching for dark matter. To test the sensitivity of this search we compare the results of such a razor analysis with 800 pb$^{-1}$ to a mono-jet analysis which uses 1 fb$^{-1}$~\cite{Fox:2011pm}, and show how the bounds from these two complementary analyses can be combined\footnote{We use  800 pb$^{-1}$ of data to match the most recent razor search, but our techniques can easily adapted to upcoming updates to this analysis. }. 

\subsection{The Razor Variables}\label{sec:variables}

The objective of the razor analysis is to discriminate the kinematics of heavy pair production from those of the SM backgrounds, without making any strong assumptions about the $\met$ spectrum or the details of the subsequent decay chains. Furthermore, background events follow very clean exponential distributions in the razor variables which allow for data-driven analyses to be carried out, without heavy use of Monte-Carlo simulations to predict backgrounds.

The baseline selection requires at least two reconstructed objects in the final state, {\emph{i.e.}} calorimetric jets or electrons and muons that satisfy lepton selection criteria. These objects are combined into two ``megajets''.  In our analysis most events contain only two jets in which case each jet is promoted to a megajet, but in the most general case the megajets are created using a ``hemisphere" algorithm described below ~\cite{HemisphereAlgo}.
The hemispheres are defined by $P_i(i=1,2)$ which is the sum of the momenta of high $p_T$ objects in the hemisphere. The high $p_T$ objects $k$ in hemisphere $i$ satisfy $d(p_k,P_i)<d(p_k,P_j)$ where   $d(p_k,P_i)\equiv(E_i-|\vec P_i|\cos\theta_{ik})\frac{E_i}{(E_i+E_k)^2}$, and $\theta_{ik}$ is the angle between $\vec P_i$ and $\vec p_k$. The hemisphere axes, $P_i$, are defined by the following algorithm.
\begin{enumerate}
\item Assign $P_1$ to the object (jet, lepton, photon) with the highest $p_T$ and $P_2$ to the object that gives the largest invariant mass as a pair with $P_1$. The four-momenta $P_1,\,P_2$ are the seeds for the hemisphere axes.
\item Go through the rest of the objects in the event, ordered by $p_T$, and assign $p_k$ to hemisphere $1$ if $d(p_k,P_1)<d(p_k,P_2)$, or $2$ otherwise. 
\item Redefine $P_i$ as the sum of the momenta in the $i^{\mathrm{th}}$ hemisphere.
\item Repeat 2-3 until all objects are assigned to a hemisphere.
\end{enumerate}
The two megajet four-momenta are taken to be the two hemisphere axes, $P_1$ and $P_2$.  

In addition to this hemisphere algorithm for defining the megajets we also considered a simple approach where the $n$ objects in an event are partitioned into two groups in all possible ($2^{n-1}-1$) ways and the partition that minimizes the sum of the megajet invariant mass-squared is chosen.  The two hemisphere algorithms give similar results. 

The razor frame is the frame in which the two megajets are equal and opposite in the $\hat z-$ (beam) direction. In this frame, the four-momenta of the megajets are
\begin{eqnarray}
p_{j_1}&=&\left(\frac{1}{2}\left[M_R-\frac{(\vec p_T^{j_1}-\vec p_T^{j_2})\cdot \vec\met}{M_R}\right],p_T^{j_1},p_z\right)\, ,\\
p_{j_2}&=&\left(\frac{1}{2}\left[M_R+\frac{(\vec p_T^{j_1}-\vec p_T^{j_2})\cdot \vec\met}{M_R}\right],p_T^{j_2},-p_z\right)\, ,
\end{eqnarray}
where  $M_R$ is the longitudinal boost invariant quantity, defined by
\begin{equation}
M_R=\sqrt{(E_{j_1}+E_{j_2})^2-(p_z^{j_1}+p_z^{j_2})^2}~.
\end{equation}
The other longitudinally invariant razor observables are
\begin{eqnarray}
M_R^T&=&\sqrt{\frac{\met(p_T^{j_1}+p_T^{j_2})-\vec{\met}\cdot(\vec{p}_T^{j_1}+\vec{p}_T^{j_2})}{2}}\, ,\\
R&=&\frac{M_R^T}{M_R}~,
\end{eqnarray}
here $p_T=|\vec{p}_T|$.  Note that the missing transverse energy, $\vec{\met}$ is calculated from all activity in the calorimeters whereas $\vec{p}_T^{j_{1,2}}$ involve just the jets above our cuts.

$M_R$ provides an estimate of the underlying scale of the event. $M_R^T$ is the transverse observable that also estimates event-by-event the value of the underlying scale. The ``razor'' variable $R^2$ is designed to reduce QCD multijet background to manageable levels. $R$ is correlated with the angle between the megajets. Events where the two mega-jets are roughly co-linear have $R^2\sim1$ while events with back-to-back megajets have small $R^2$.
In general $R^2$ has a maximum value of approximately 1, and the QCD multijet background peaks at $R^2=0$. Thus, by imposing a cut on $R^2$, one can essentially eliminate the QCD multijet background. 

\subsection{Analysis}

The razor analysis uses a set of dedicated triggers which allow one to apply low thresholds on $M_R$ and $R^2$.  The events that pass the triggers are then classified into six disjoint boxes which correspond to different lepton selection criteria~\cite{CMSEWK}.  For our purposes, we consider only the HAD box which contains all the events that fail lepton requirements, described below. After QCD is removed using a strong $R^2$ cut, the dominant backgrounds to our process are $(Z\to \bar\nu\nu)+$jets, $(W\to \ell^{inv}\nu)+$jets, $(W\to \tau^{h}\nu)+$jets, and $t\bar t$, where $\ell^{inv}$ denotes a lepton that is missed in the reconstruction, and $\tau^h$ is a hadronically decaying tau-lepton.  We have simulated the dominant SM backgrounds using MadGraph5 \cite{Alwall:2011uj} at the matrix element level, Pythia 6.4 \cite{Sjostrand:2006za} for parton showering and hadronization, and PGS \cite{PGS} as a fast detector simulation. We generate $W/Z+n$ jets, where $n=1,2,3$ for the background, and use MLM matching \cite{Hoche:2006ph} with a matching scale of 60 GeV.  We generate both matched and unmatched samples for our signal, and find that the matched sample gives approximately a 15\% increase in the number of events passing our analysis cuts, as compared to the unmatched sample. In what follows, we use unmatched samples for the signal events; using a matched sample will increase our bounds by a few GeV but does not change our conclusions.  The cross sections for the dominant backgrounds, and an example signal point, are shown in Table~\ref{tab:xsecs}.

\begin{table}
   \centering
   \parbox{14cm}{
   \begin{ruledtabular}
   \begin{tabular}{cccccc} 
             & $n_j=0$ & $n_j=1$ & $n_j=2$ & $n_j=3$ & After cuts \\
    \hline
 $(Z\to \bar\nu\nu)+$jets & 3960  & 470  & 150  & 33.7  & $18 \times 10^{-2}$ \\
 $(W\to \ell^{inv}\nu)+$jets & 10585  & 836  & 317  & 96.5  & $2.0 \times 10^{-2}$ \\ 
 $(W\to \tau^{h}\nu)+$jets & 5245 & 676  & 160  & 48.8  &  $6.8 \times 10^{-2}$ \\ 
 $t\bar{t}$ & 12.4  & --  & --  & --  & $1.5 \times 10^{-3}$ \\     
 $\bar{\chi}\chi$ & 5.46 & 2.31  & 0.77  & 0.33  & $4.3 \times 10^{-2}$ \\ 
   \end{tabular}
   \end{ruledtabular}}
   \caption{Background and signal (for $m_\chi=100$ GeV and $\Lambda= 644$ GeV) cross sections (in pb) before and after analysis cuts. $n_j$ is the number of jets.  The matching scale is taken to be 60 GeV, see text for details.}
   \label{tab:xsecs}
\end{table}

Following~\cite{CMSrazor}, in every event we require jets to have $p_T>60$ GeV, $|\eta|<3.0$.  Electrons(muons) are required to have $p_T>20 (10)$ GeV and $|\eta|<2.5(2.1)$, and we include $\tau$-leptons, which decay hadronically, in our definition of jets.   Only events in which $\Delta\phi$ between the two megajets is less than 2.8 are kept.
With these requirements the events will pass the dedicated razor triggers, although they would often fail those for other analyses {\emph{e.g.}} $\alpha_T, H_T$.  One advantage of the razor analysis lies in the simple shape of the SM background distributions; the $M_R$ and $R^2$ distributions are simple exponentials for a large portion of the $R^2-M_R$ plane. By fitting the distributions of the razor variables $M_R$ and $R^2$ to an exponential function, one can utilize a data-driven description of the background without having to rely on Monte Carlo (MC) estimates. Since we do not have access to the data, we must carry out a MC based analysis.  As a check of the validity of our MC analysis we compare our results to the yields found by CMS in different bins of $R^2$ and $M_R$.  We find that our MC simulations for the background in the HAD box fall within the expected 68\% range expected by CMS, and thus are consistent with the CMS simulations (see Fig. 9 of Ref.~\cite{CMSrazor}), which in turn agree well with data.

\subsection{Signal and Background Shapes}\label{sec:results}

\begin{figure}[t]
\centering
\subfigure[~$(Z\to \bar\nu\nu)+$jets.]{\includegraphics[width=0.42\textwidth, height=0.41\textwidth]{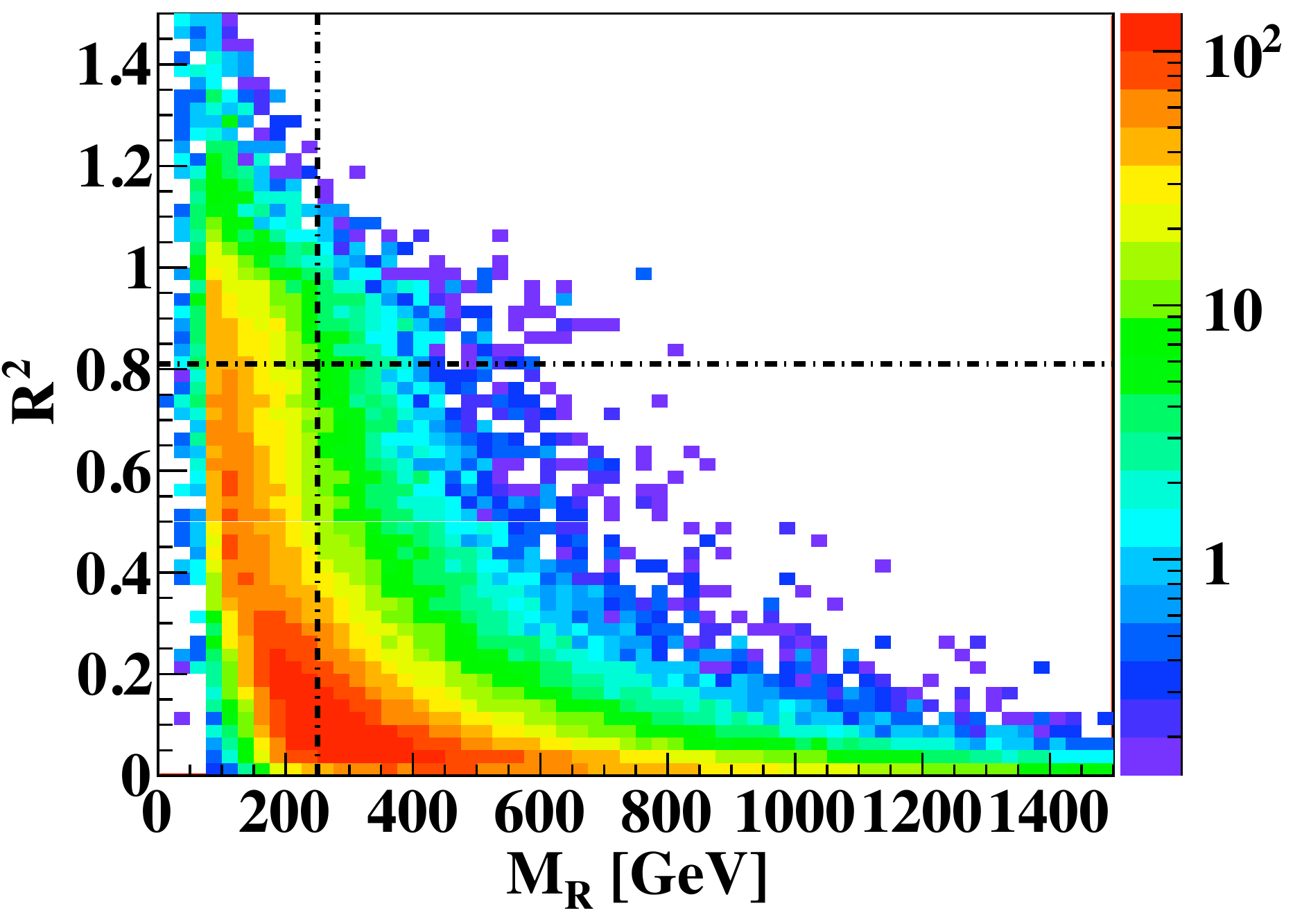}
\label{fig:rmr_zjets}}
\subfigure[~$W+$jets.]{\includegraphics[width=0.42\textwidth,height=0.42\textwidth]{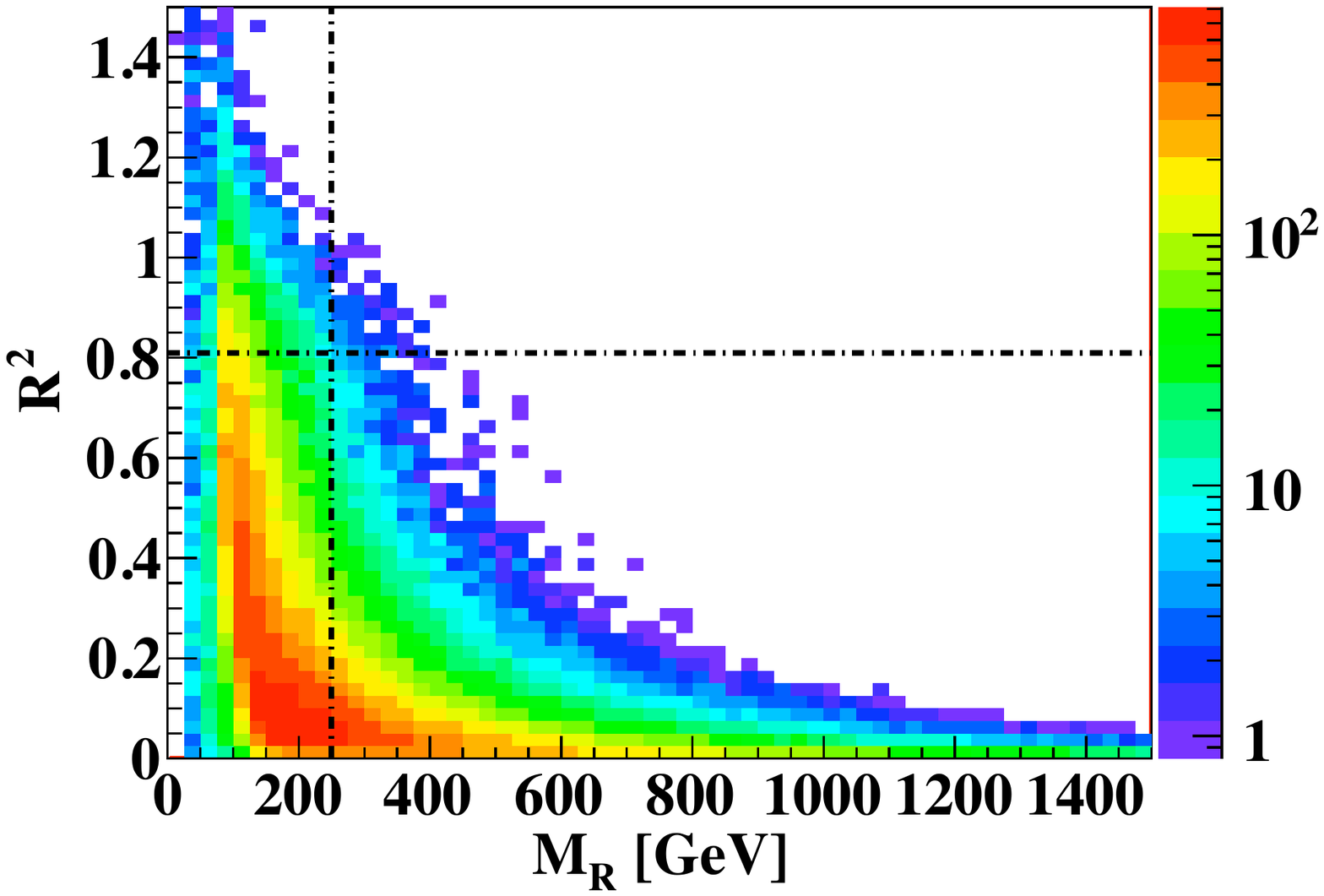}
\label{fig:rmr_wjets}}\\
\subfigure[~$t\bar t$.]{\includegraphics[width=0.42\textwidth,height=0.42\textwidth]{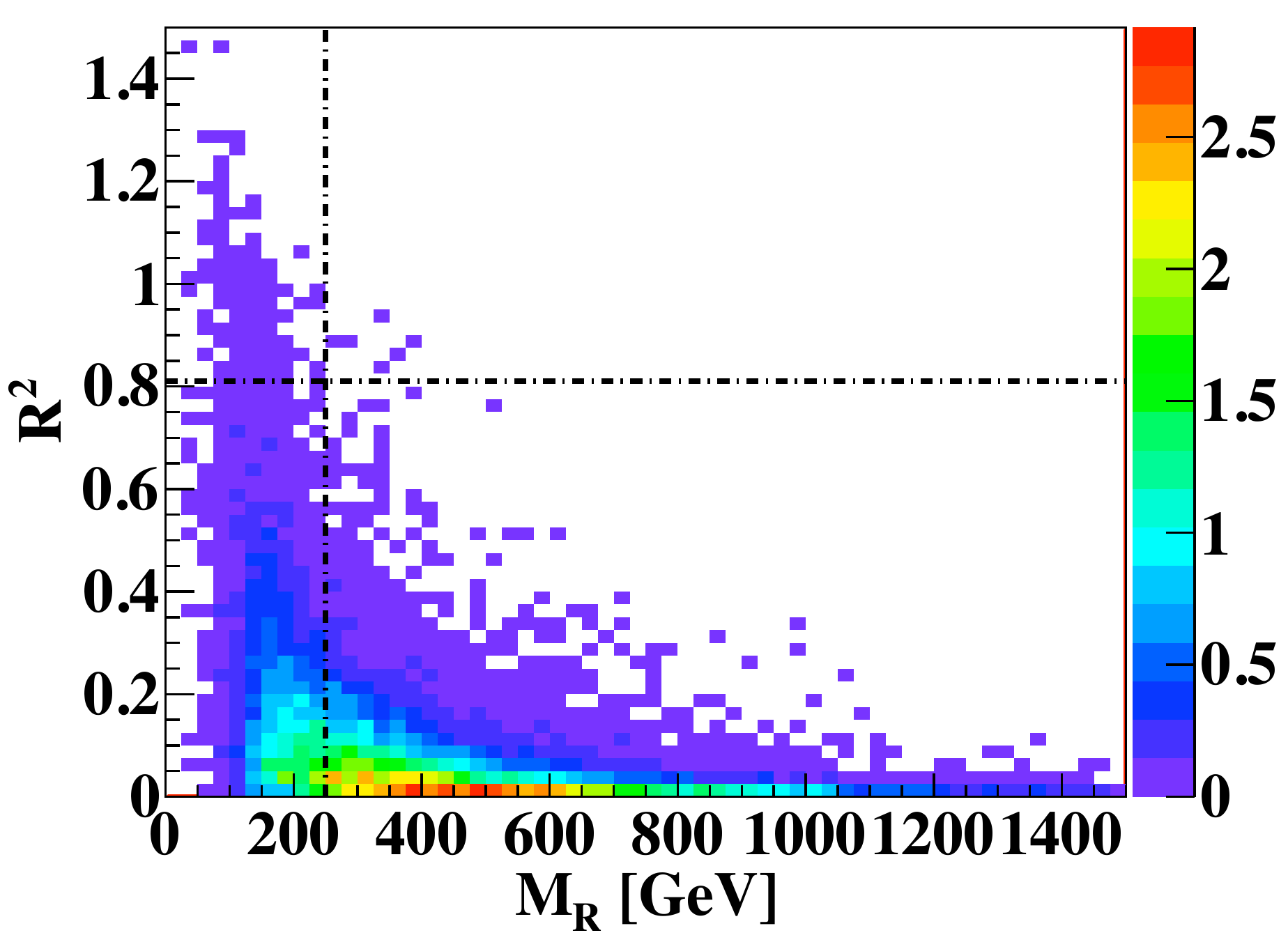}
\label{fig:rmr_ttbar}}
\subfigure[~Signal ($M_{\chi}=100$ GeV, $\Lambda = 644$ GeV).]{\includegraphics[width=0.42\textwidth,height=0.42\textwidth]{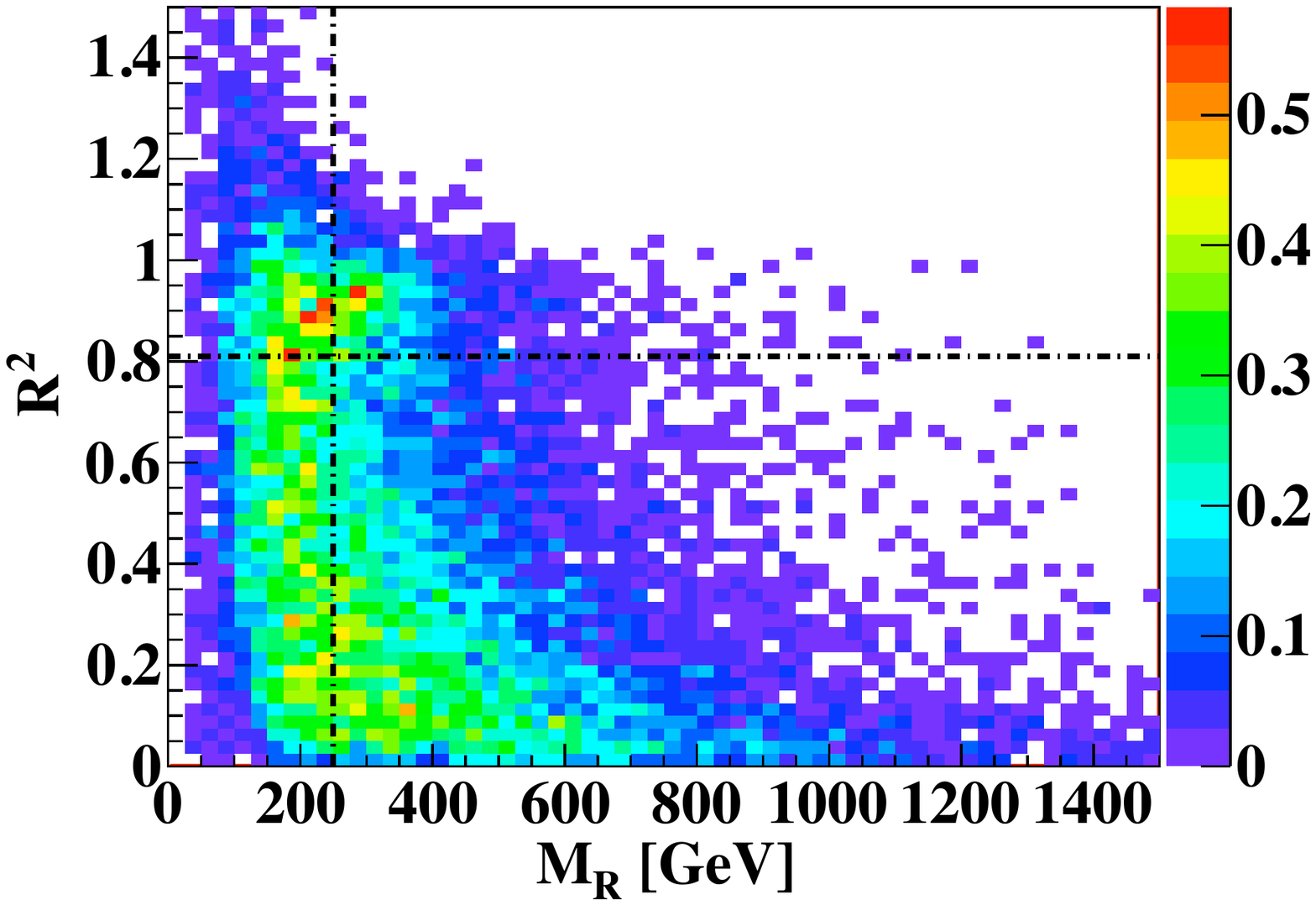}
\label{fig:rmr_dm}}
\caption{$R^2$ vs. $M_R$ distribution for SM backgrounds (a) $(Z\to \bar\nu\nu)+$jets, (b) $W+$jets (including decays to both $\ell^{inv}$ and $\tau^h$, (c) $t\bar t$, and (d) DM signal with $M_{\chi}=100$ GeV and $\Lambda= 644$ GeV.  In all cases the number of events are what is expected after an integrated luminosity of 800 pb$^{-1}$.  The cuts applied in $M_R$ and $R^2$ are shown by the dashed lines and the ``signal" region is the upper right rectangle.}
\label{fig:rmr_compare}
\end{figure}
%
\begin{figure}[th]
\begin{center}
\subfigure[~$M_{\chi}=0.01$ GeV.]{\includegraphics[width=0.42\textwidth, height=0.42\textwidth]{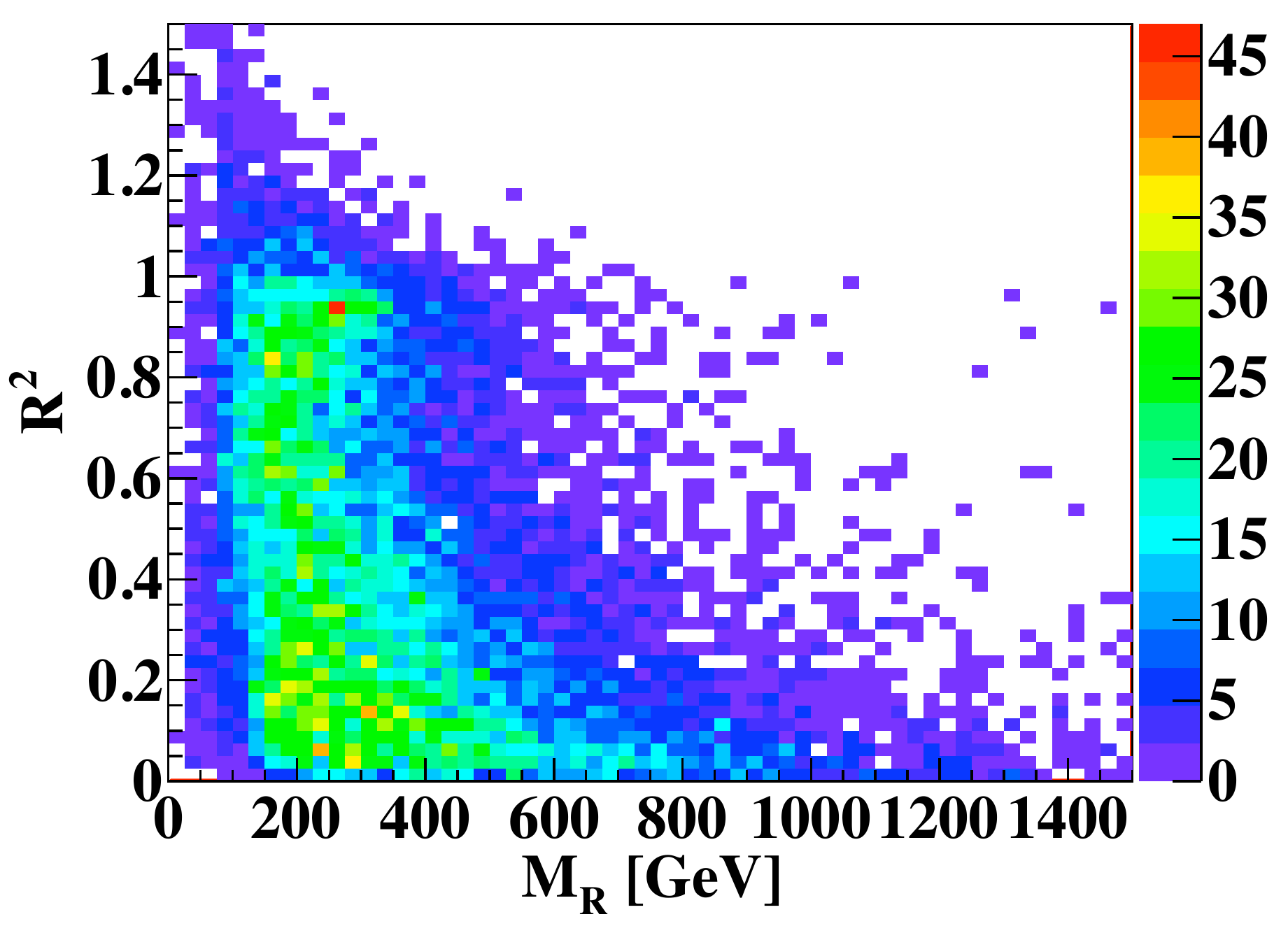}
\label{fig:DM_rmr_1}}
\subfigure[~$M_{\chi}=100$ GeV.]{\includegraphics[width=0.42\textwidth, height=0.42\textwidth]{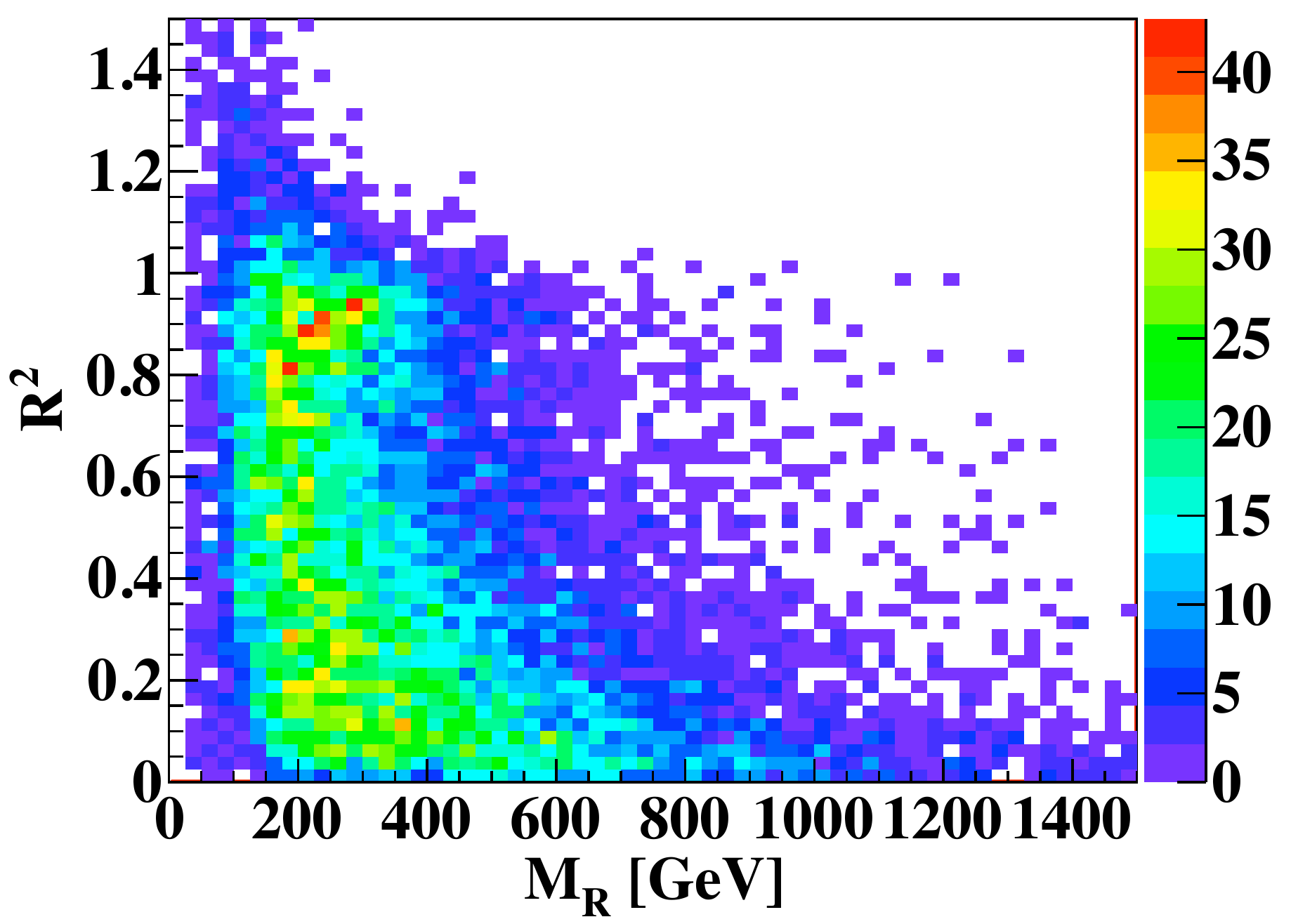}
\label{fig:DM_rmr_5}}\\
\subfigure[~$M_{\chi}=800$ GeV.]{\includegraphics[width=0.42\textwidth, height=0.42\textwidth]{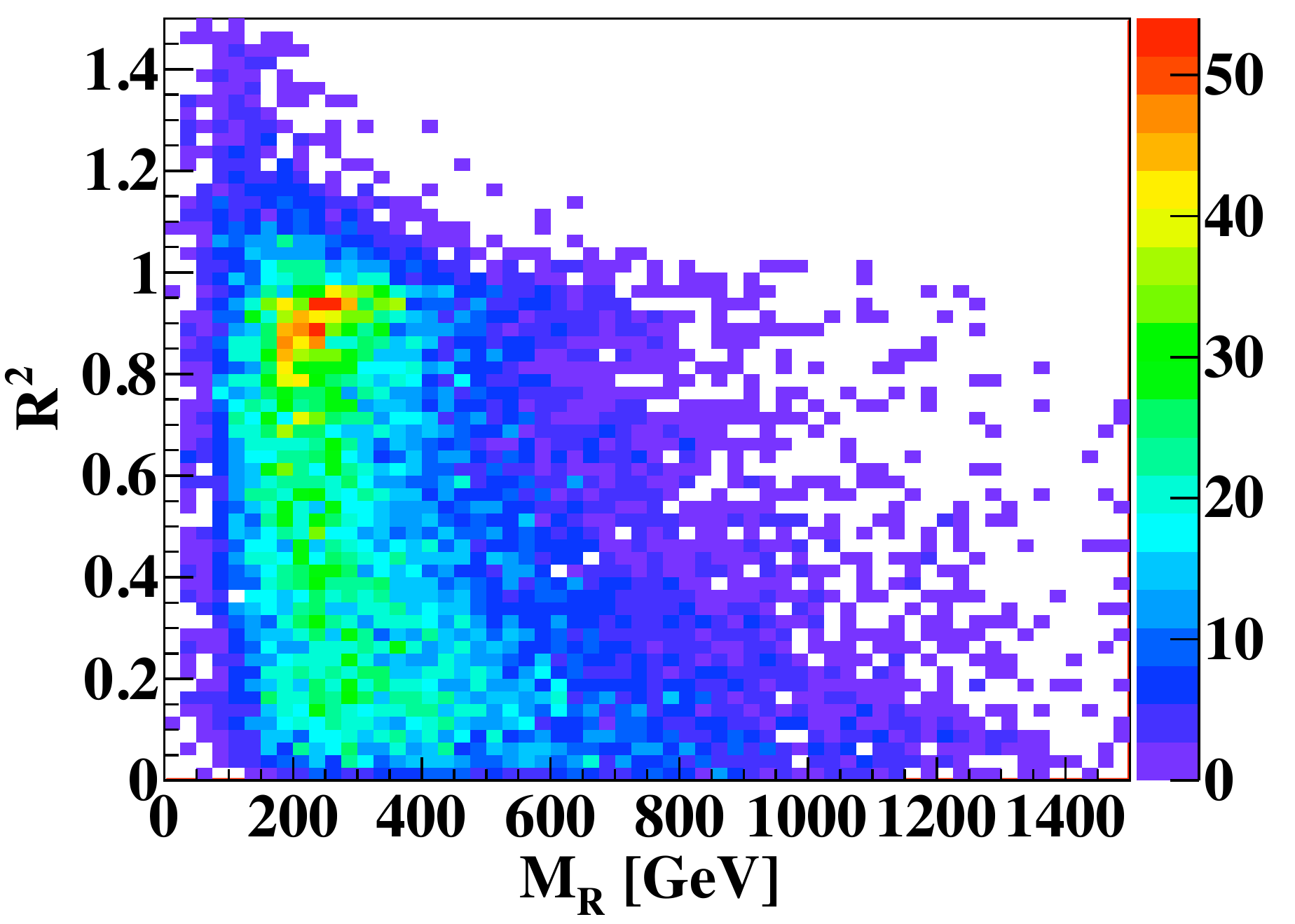}
\label{fig:DM_rmr_12}}
\subfigure[~$M_{\chi}=1000$ GeV.]{\includegraphics[width=0.42\textwidth, height=0.42\textwidth]{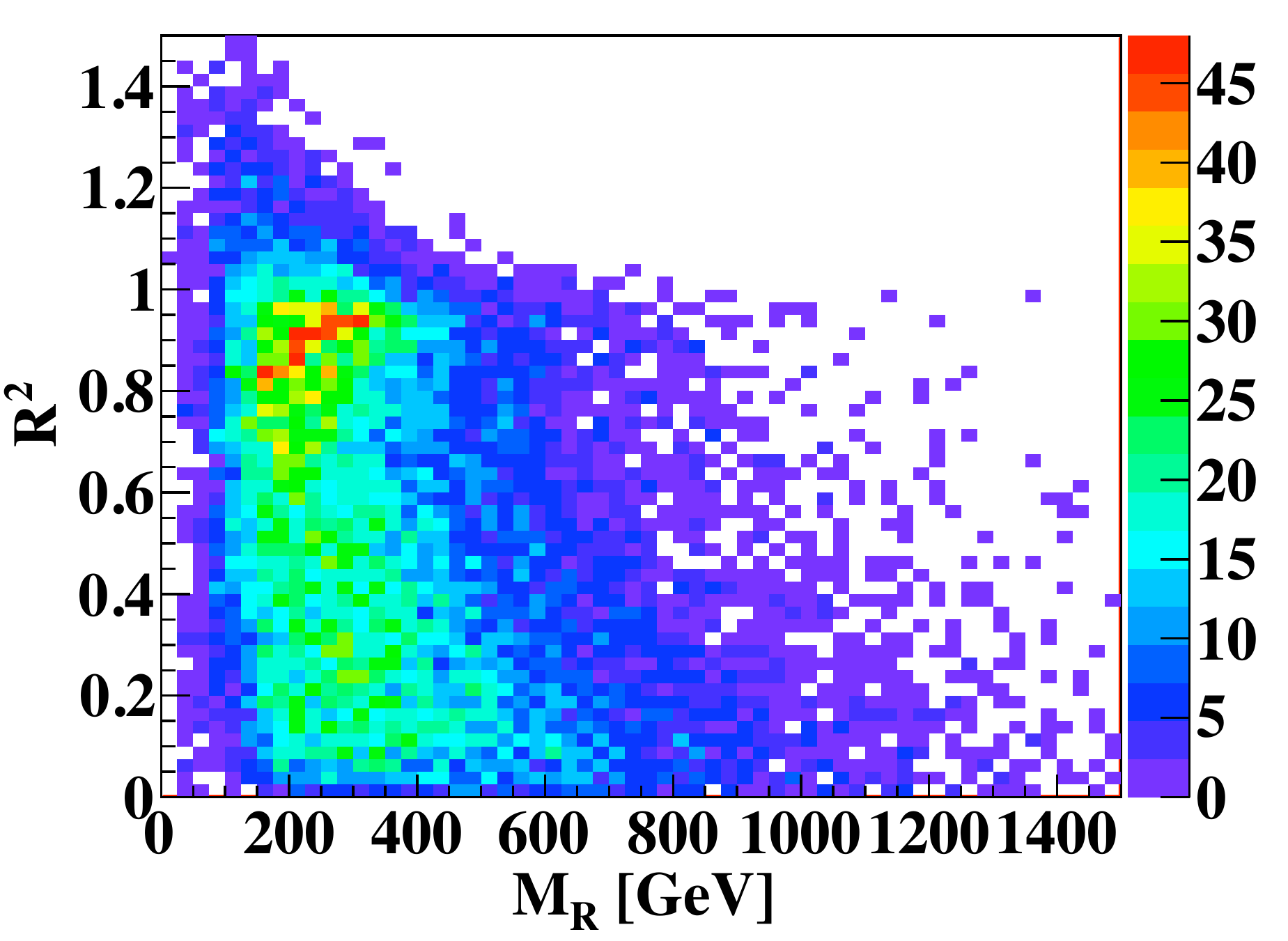}
\label{fig:DM_rmr_14}}
\caption{$R^2$ vs. $M_R$ for various DM masses with $u$-only vectorial couplings with arbitrary normalization.}
\label{fig:DM_rmr}
\end{center}
\end{figure}
\begin{figure}[t]
\begin{center}
\subfigure[~Sea quark couplings.]{\includegraphics[width=0.42\textwidth, height=0.42\textwidth]{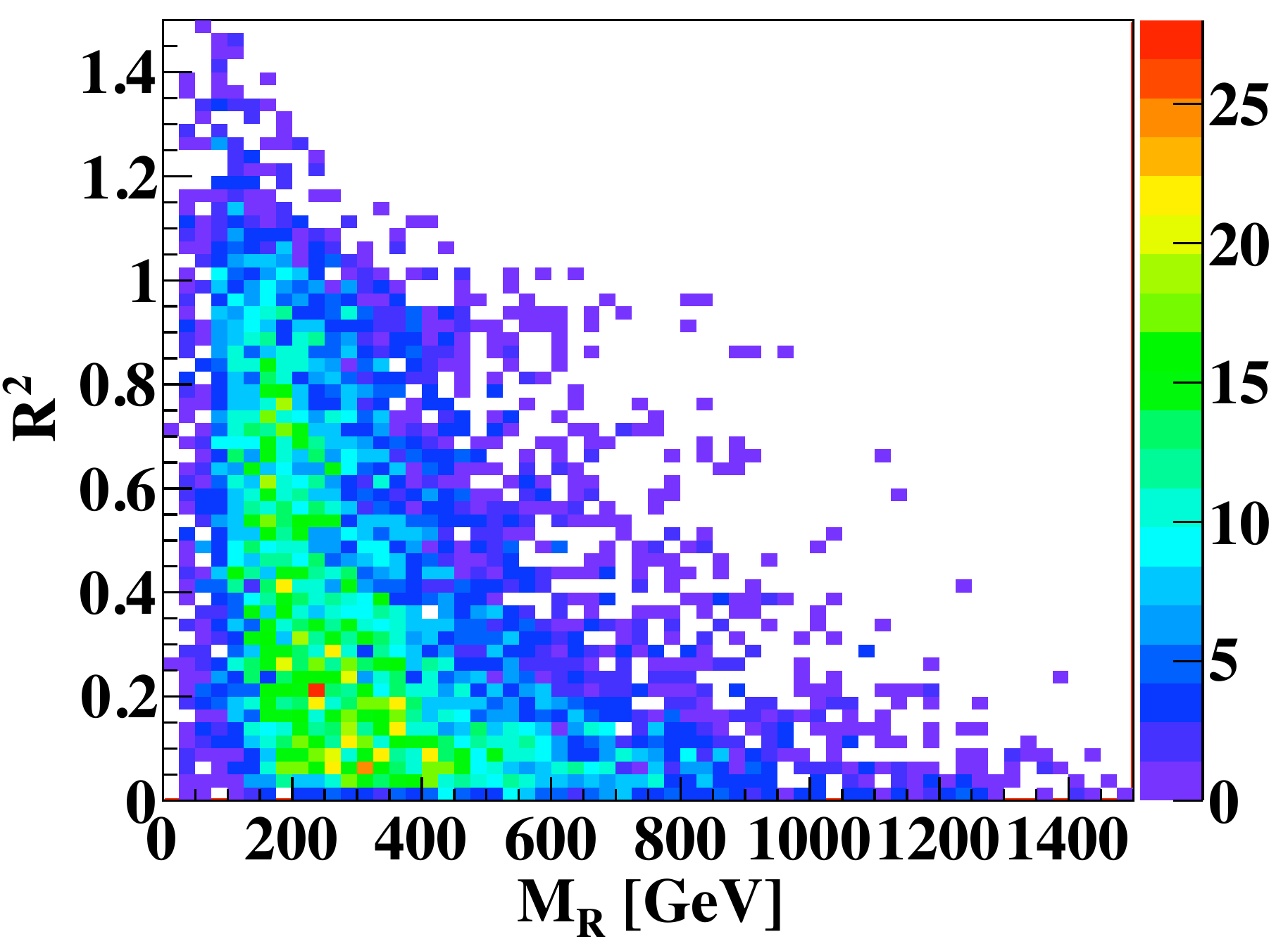}
\label{fig:sea_rmr}}
\subfigure[~Gluon couplings.]{\includegraphics[width=0.42\textwidth, height=0.42\textwidth]{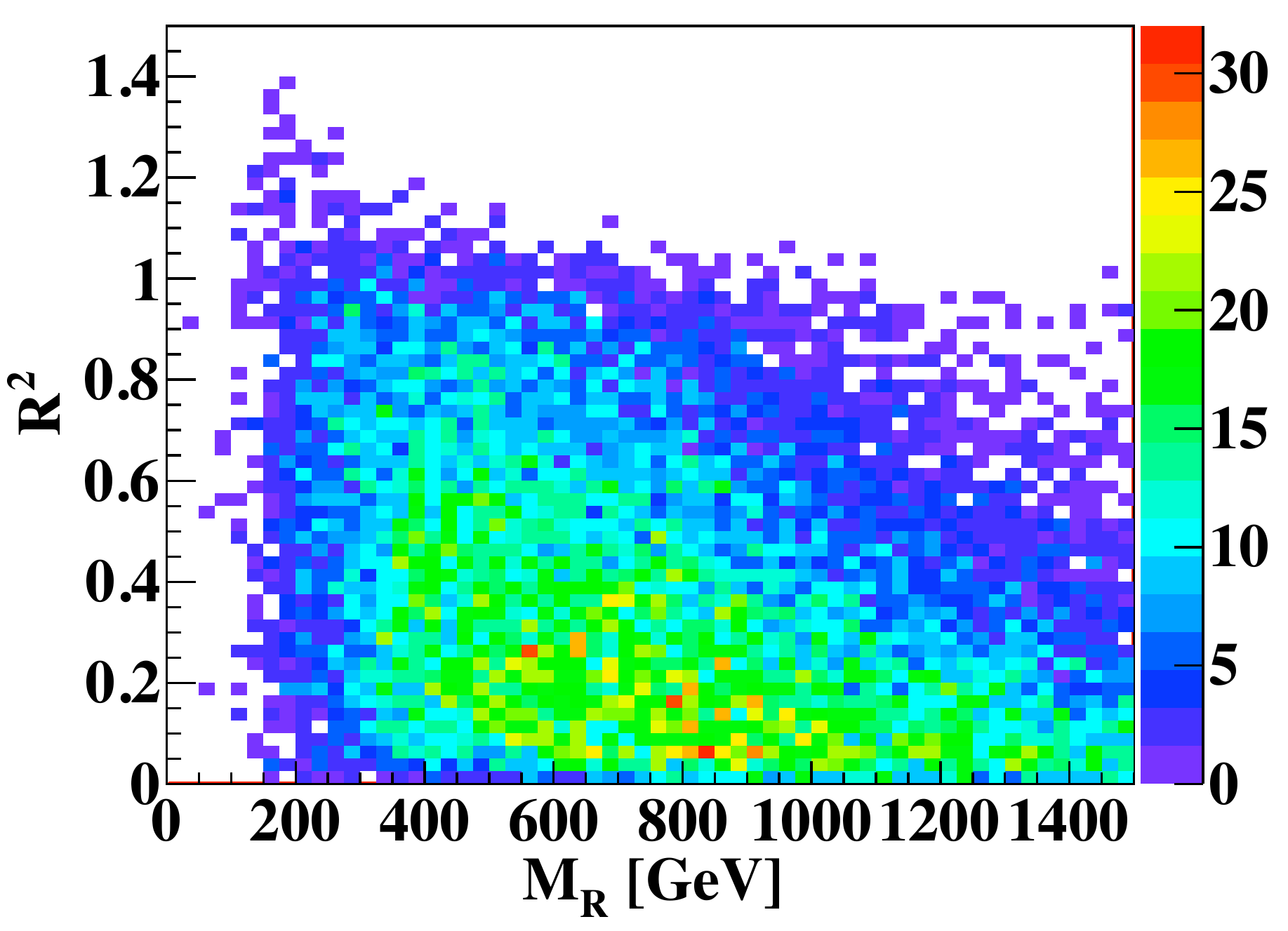}
\label{fig:gluon_rmr}}
\caption{$R^2$ vs. $M_R$ for DM coupling to (a) sea quarks (in this case the $s$-quark) and (b) gluons with arbitrary normalization.}
\label{fig:alt_rmr}
\end{center}
\end{figure}

The shape of the $M_R$ and $R^2$ distributions for the dominant backgrounds and a sample signal are shown in Fig.~\ref{fig:rmr_compare}. The dependence of the signal shape on dark matter mass is shown in Fig.~\ref{fig:DM_rmr}. The signal shapes when dark matter couples to sea quarks or to gluons are shown in Fig.~\ref{fig:alt_rmr}.
The shapes depend on the scale and the kinematics of the production process.
The location of the $M_R$ distribution peak is determined by the event scale and kinematic cuts. The $M_R$ distributions of $(Z\to \bar\nu\nu)+$jets, $W+$jets, and $\bar\chi\chi+$jets all peak at approximately the same value of $M_R\approx 200$ GeV, whereas the $M_R$ peak for $t\bar t$ is higher due to the inclusion of tops in the megajets. 

The shape of $R^2$ distribution is affected by the kinematics of the process and is somewhat different for signal and background. Background events are highly peaked at low $R^2$, where the megajets are more back-to-back, whereas signal events are more evenly distributed in $R^2$, with a significant population at high $R^2$. The difference in event shapes, signal events being more likely to produce collinear megajets, originated from different diagrams which dominate production.

The SM backgrounds are dominated by invisible decays of a $Z$ boson, see Table~\ref{tab:xsecs}, for which the dominant production mechanism at the LHC is through quark-gluon collisions with $q\bar q$ collisions giving a much smaller contribution. In quark-gluon collisions the $Z$ tends to be emitted in the backward direction (close to the beam from which the gluon came). This tends to give the $Z$ a lower $p_T$ compared to events which originate in $q\bar{q}$. Due to the high $p_T$ cuts on the individual jets their transverse momenta must largely cancel to balance the $Z$.  Thus, the $\Delta\phi$ distribution is peaked near $\pi$ for background.  

On the other hand, signal events are dominantly produced from the $q\bar{q}$ initial state. This is because $q\bar{q}$ and $qg$ initiated cross sections scale differently with the invariant mass of the dark matter pair. 
This is reminiscent of the scaling of $Z+j$ at LHC, where the $gq$-initiated cross section is proportional to $m_Z^2$ while the $q\bar{q}$-initiated one scales like $m_Z^4$. If the $Z$ mass were higher, $Z+j$ would have been dominantly $q\bar{q}$-initiated. Similarly  in our case DM production is dominatly $q\bar{q}$-initiated because the $\chi\bar{\chi}$ invariant mass (analogous to the $Z$ mass above) is typically far above the weak scale, see Figure~\ref{fig:mchichi}.  
This difference in production mechanisms results in a more isotropic distribution of the jets and consequently a different distribution in $R^2$, tending more towards high values.  This difference increases as DM mass increases, as the peak in $R^2$ also moves higher as DM mass increases (Fig.~\ref{fig:DM_rmr}) while the $M_R$ distribution remains approximately the same. The difference in production mechanisms remains at NLO, which we have checked using MCFM ~\cite{Campbell:2011bn,MCFM}. 

We also find that the $M_R$ and $R^2$ distributions for DM coupling to sea quarks, shown in Fig.~\ref{fig:alt_rmr}, are similar to those of background.  This is because for sea quarks the dominant production is $qg$ (as well as $\bar qg$) because of their smaller PDF's, which is similar to the dominant background production mechanism. For coupling to gluons, where the $gg$ initial state dominates, the distribution gives a more even coverage of the $M_R - R^2$ plane, as seen in Fig.~\ref{fig:alt_rmr}.

\subsection{Results}

\begin{figure}[t]
\begin{center}
\subfigure[~Vector couplings.]{\includegraphics[width=0.48\textwidth]{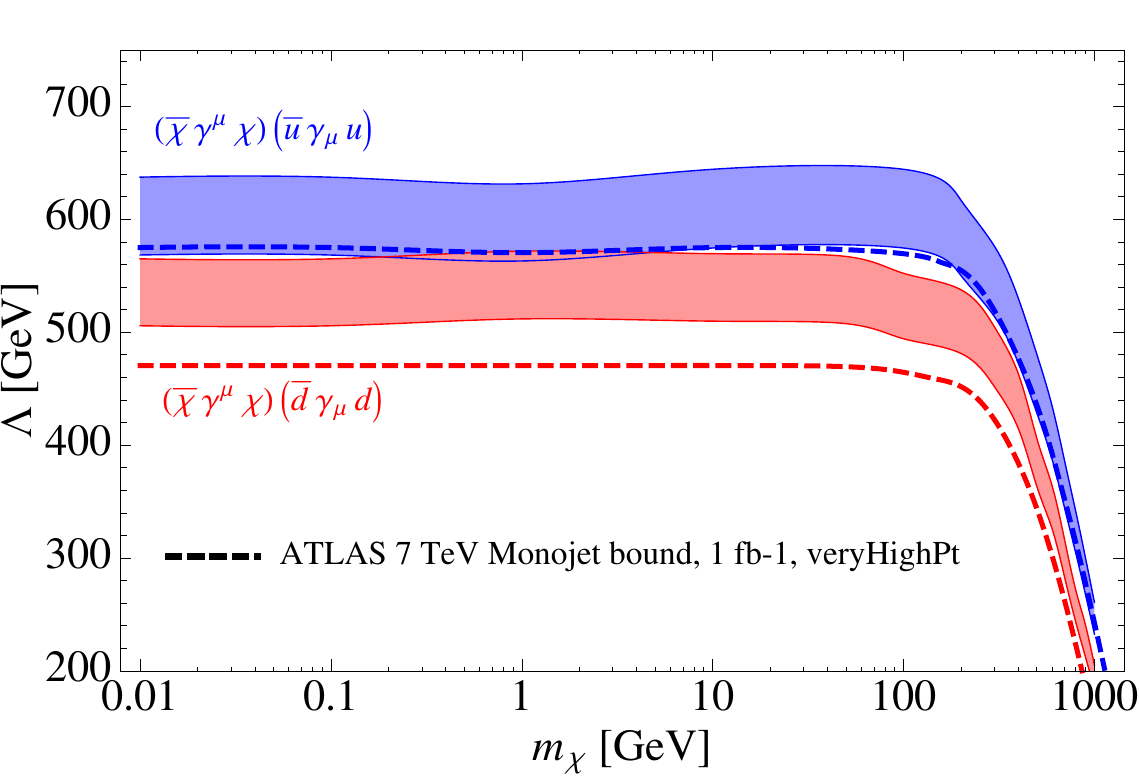}
\label{fig:DM_vbounds}}
\subfigure[~Axial couplings]{\includegraphics[width=0.48\textwidth]{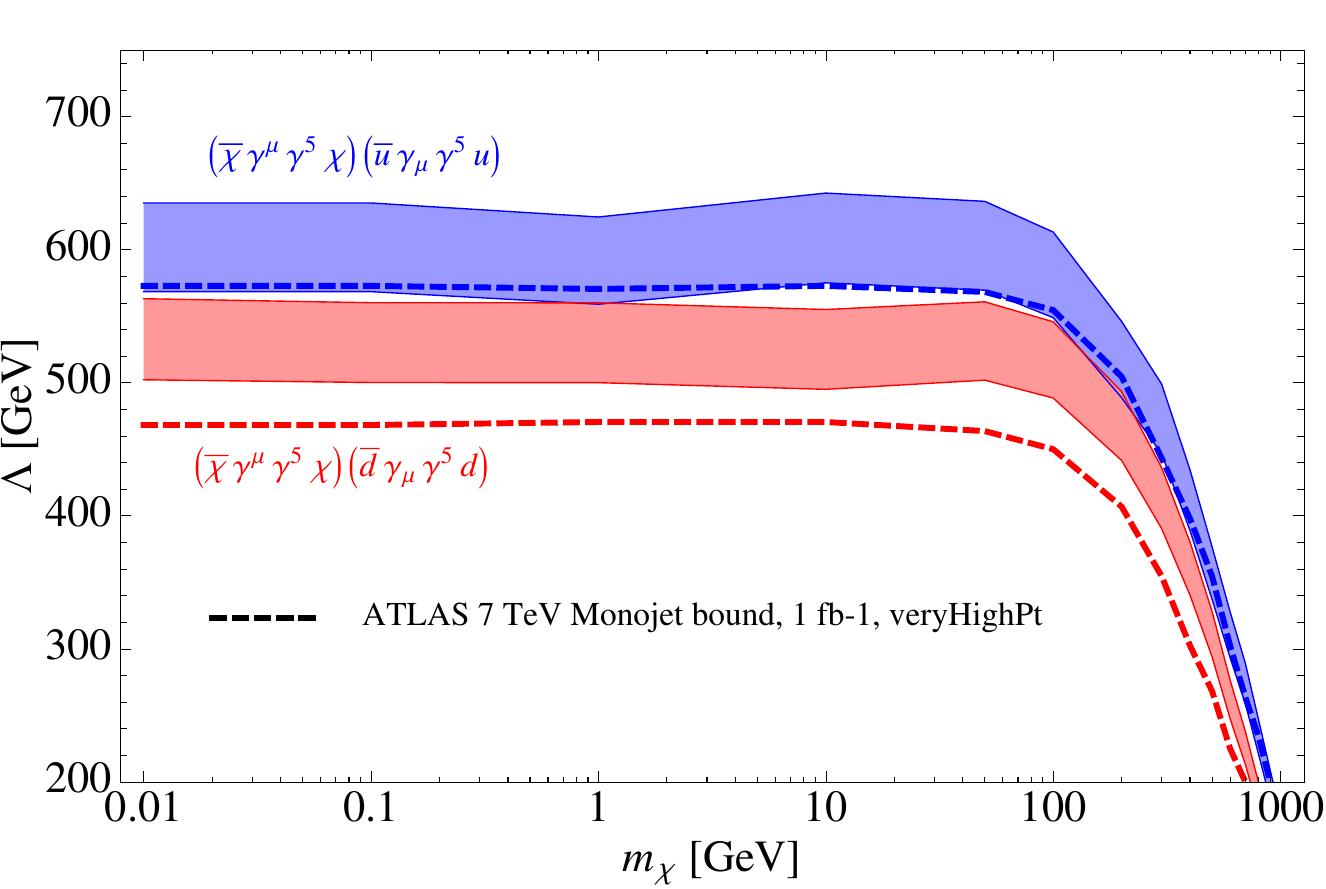}
\label{fig:DM_abounds}}
\subfigure[~Gluon couplings]{\includegraphics[width=0.48\textwidth]{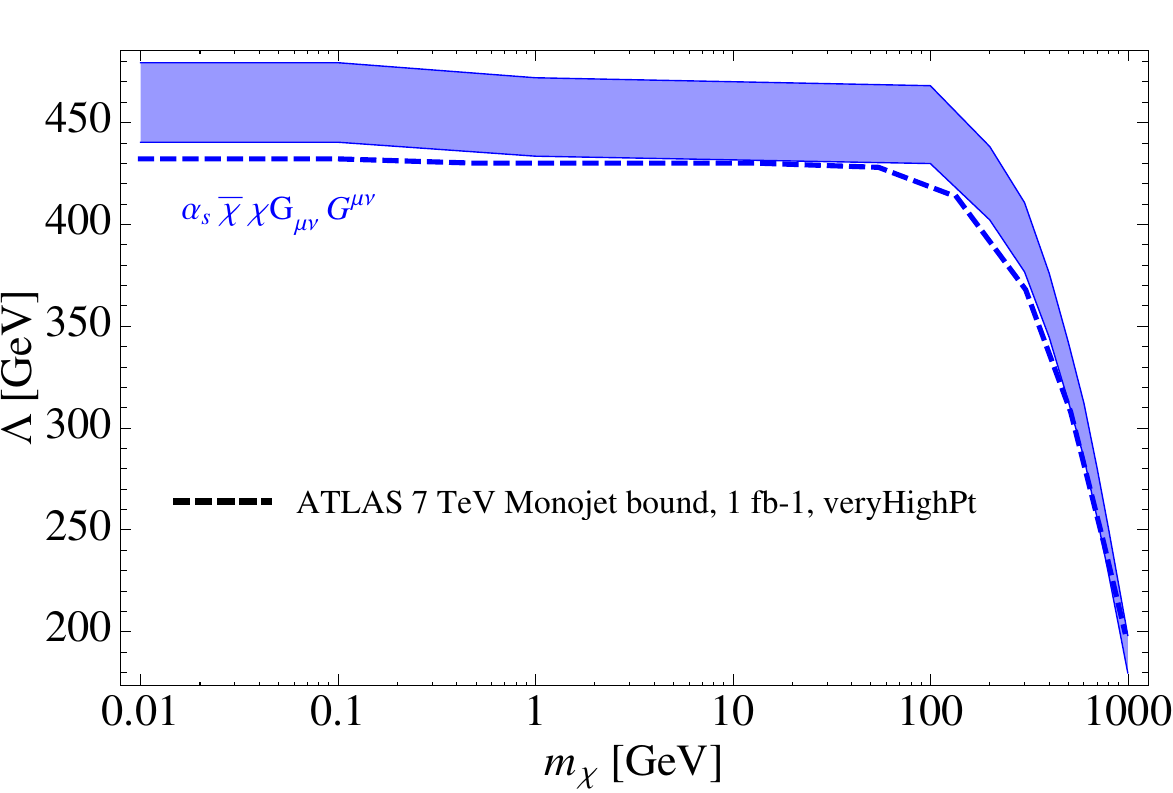}
\label{fig:DM_gbounds}}
\caption{Cutoff scale $\Lambda$ bounds for vector, axial, and gluon couplings. The error band is determined by varying $\sigma_{SM}$ between $\sqrt{N_{SM}}$ and $\sigma_{SM} = 2\sqrt{N_{SM}}$. The dashed line is the bound determined by the monojet analysis \cite{Fox:2011pm}.}
\label{fig:DM_bounds}
\end{center}
\end{figure}

Based on the distributions shown in Figs.~\ref{fig:rmr_compare},~\ref{fig:DM_rmr}, and ~\ref{fig:alt_rmr}, we find that our optimal signal region is $M_R \geq 250\  \textrm{GeV}$ and $R^2 \geq 0.81$. We use the number of events in the signal region, the upper right rectangle in Fig.~\ref{fig:rmr_compare}, to place estimated constraints on the cutoff scale $\Lambda$.  At $90\%$ exclusion, we require 
\begin{equation}
\chi^2\equiv \frac{N_{DM}(m_{\chi},\Lambda)^2}{N_{DM}(m_{\chi},\Lambda)+N_{SM}+\sigma_{SM}^2}\le2.71\, ,
\label{eq:chi2}
\end{equation}
where $N_{DM}$ is the expected number of signal events for a given DM mass $m_{\chi}$ and scale $\Lambda$, $N_{SM}$ is the expected number of background events, and $\sigma_{SM}$ is the uncertainty in the predicted number of background events. Through our Monte Carlo simulations, we estimate that the number of background events is $144.0$ for $(Z\to \bar\nu\nu)+$jets, $70.4$ for $W+$jets, and $1.2$ for $t\bar t$, giving a total of $N_{SM}=215.6$ for a luminosity of $800 \textrm{ pb}^{-1}$, the approximate amount used in the Razor analysis~\cite{CMSrazor}. The $t\bar t$ background does not give a large contribution since the majority of events with significant $\met$ are vetoed by the presence of leptons in the events and do not pass our cuts. We did not attempt to calculate the QCD background since we expect a negligible number of events from this channel in our signal region. The error $\sigma_{SM}$ in the razor analysis is statistics dominated which implies $\sigma_{SM}\sim\sqrt{N_{SM}}$. We adopt this value as our default value for the standard model uncertainty, but to be conservative we will also present the limit in the case where there is an additional and equal source of systematic error. 
The calculated bound for vector and axial couplings of DM to valence quarks is given in Fig.~\ref{fig:DM_bounds}, where we see that the existing razor analysis gives bounds that are competitive with the monojet results. We present the limit as a band extending between the two assumptions for the uncertainty $\sigma_{SM} = \sqrt{N_{SM}}$ and $\sigma_{SM} = 2\sqrt{N_{SM}}$. In the rest of the paper we use the $\sqrt{N_{SM}}$ limit which we expect to be realistic. Note that, there is no significant difference between the bounds for vector or axial couplings. This implies that as opposed to direct detection, spin dependent limits will be just as strong as spin independent ones. 

\begin{figure}[t]
\begin{center}
\subfigure[~Vector-coupling]{\includegraphics[width=0.48\textwidth]{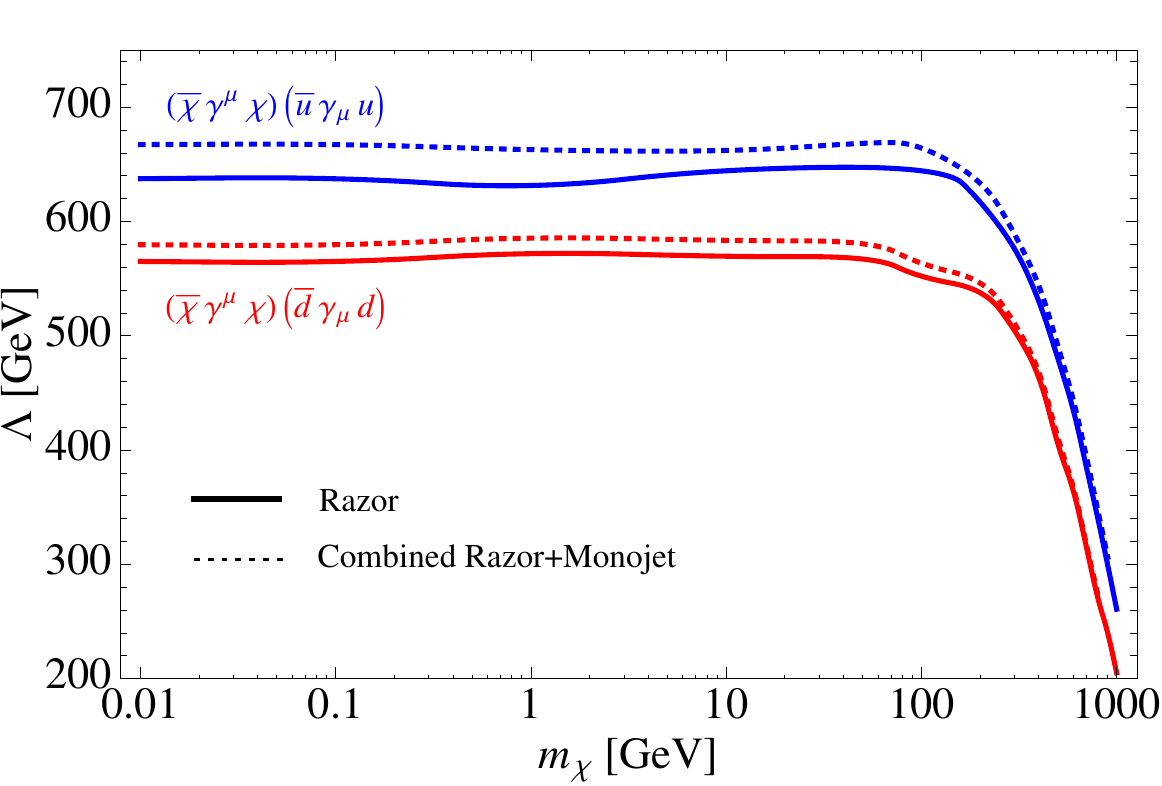}}
\subfigure[~Axial-coupling]{\includegraphics[width=0.48\textwidth]{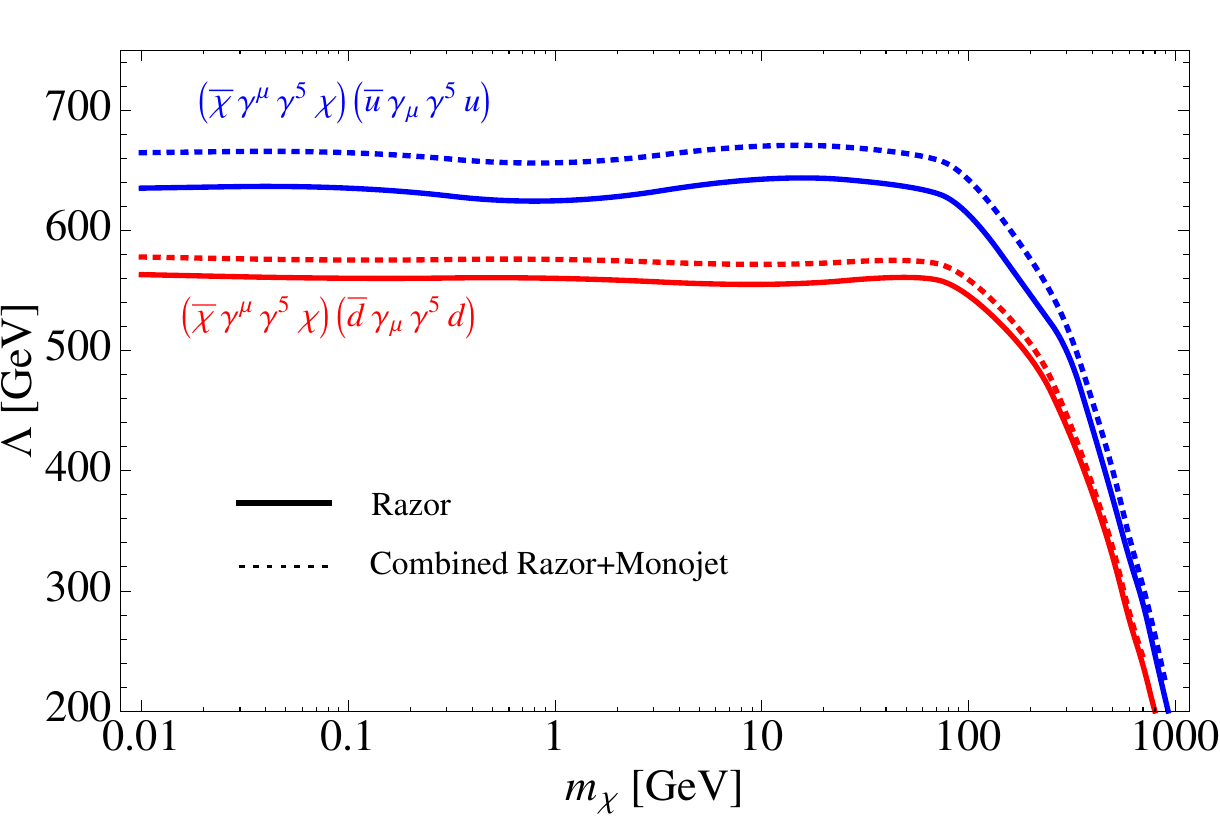}}
\subfigure[~Gluon-coupling]{\includegraphics[width=0.48\textwidth]{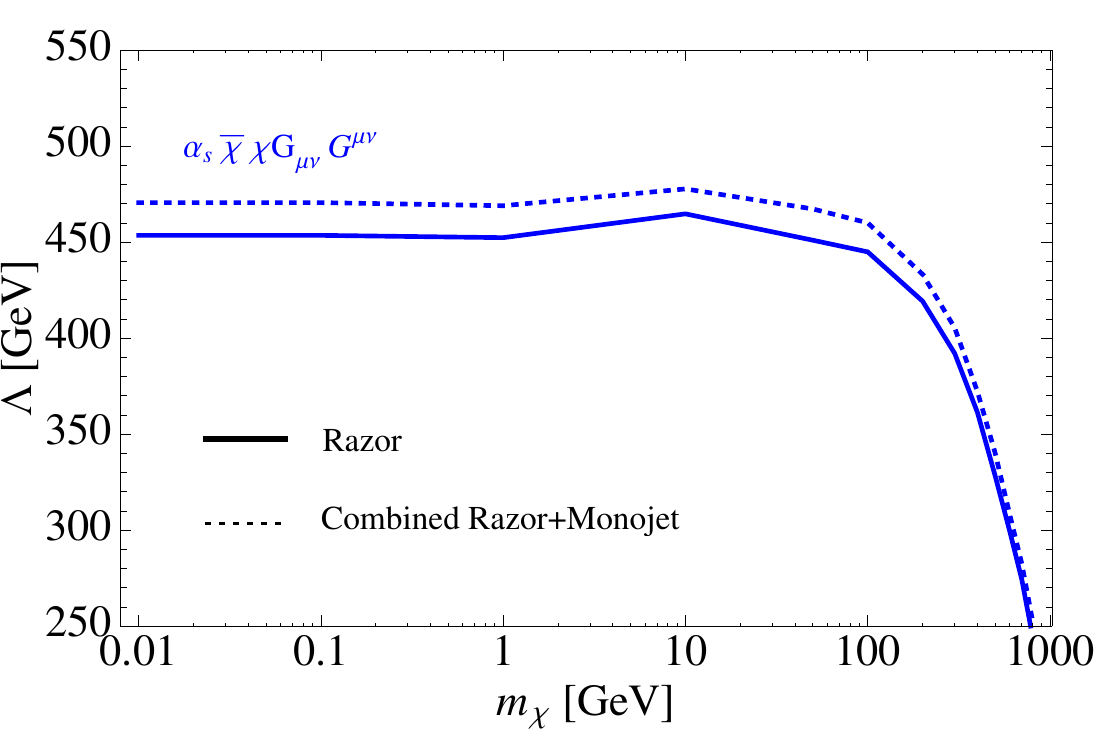}}
\caption{Combined razor and monojet $\Lambda$ bounds. The solid lines are the razor bounds and the dashed lines are the combined bounds.}
\label{fig:combined}
\end{center}
\end{figure}

The razor analysis requires at least two jets in the final state, so the data set is complementary to that used in the monojet search.  Since the bounds are slightly, but not hugely, stronger than those from monojet there is utility in combining the bounds from the razor and monojet analyses. We do this by solving 
\begin{equation}
\chi^2_{\textrm{monojet}}(m_\chi,\Lambda)+\chi^2_{\textrm{razor}}(m_\chi,\Lambda)=2.71\, ,
\end{equation}
where the $\chi^2$ are defined in Eq.~\ref{eq:chi2}. We find that the combined bound is a few percent higher than the razor bound alone (Fig.~\ref{fig:combined}).

%
\subsection{Comparison with Direct Detection and Annihilation Cross Section}\label{sec:directdetection}
We now translate the collider bounds found above into constraints on direct detection scattering rates by following the approach of Ref. \cite{Fox:2011pm}. This allows us to show the collider limits in the standard $\sigma-m_\chi$ plane. We use the values found in \cite{Bai:2010hh} to calculate the coefficients required to translate the quark level matrix elements $\langle N|\bar q\gamma^{\mu}q|N\rangle$ and $\langle N|\bar q\gamma^{\mu}\gamma^5q|N\rangle$ into the nucleon level matrix elements. For the matrix element of the gluon field strength in the nucleon, $\langle N|\alpha_sG^a_{\mu\nu}G^{a\mu\nu}|N\rangle = -\frac{8\pi}{9}\left(m_N-\sum_{q=u,d,s}\langle N|m_q\bar q q|N\rangle\right)$, we follow the approach of \cite{Ellis:2008hf} using an updated value of the pion-nucleon sigma term $\Sigma_{\pi N}= 55$ MeV \cite{Young:2009ps}. 

\begin{figure}[t]
\centering
\includegraphics[width=0.48\textwidth]{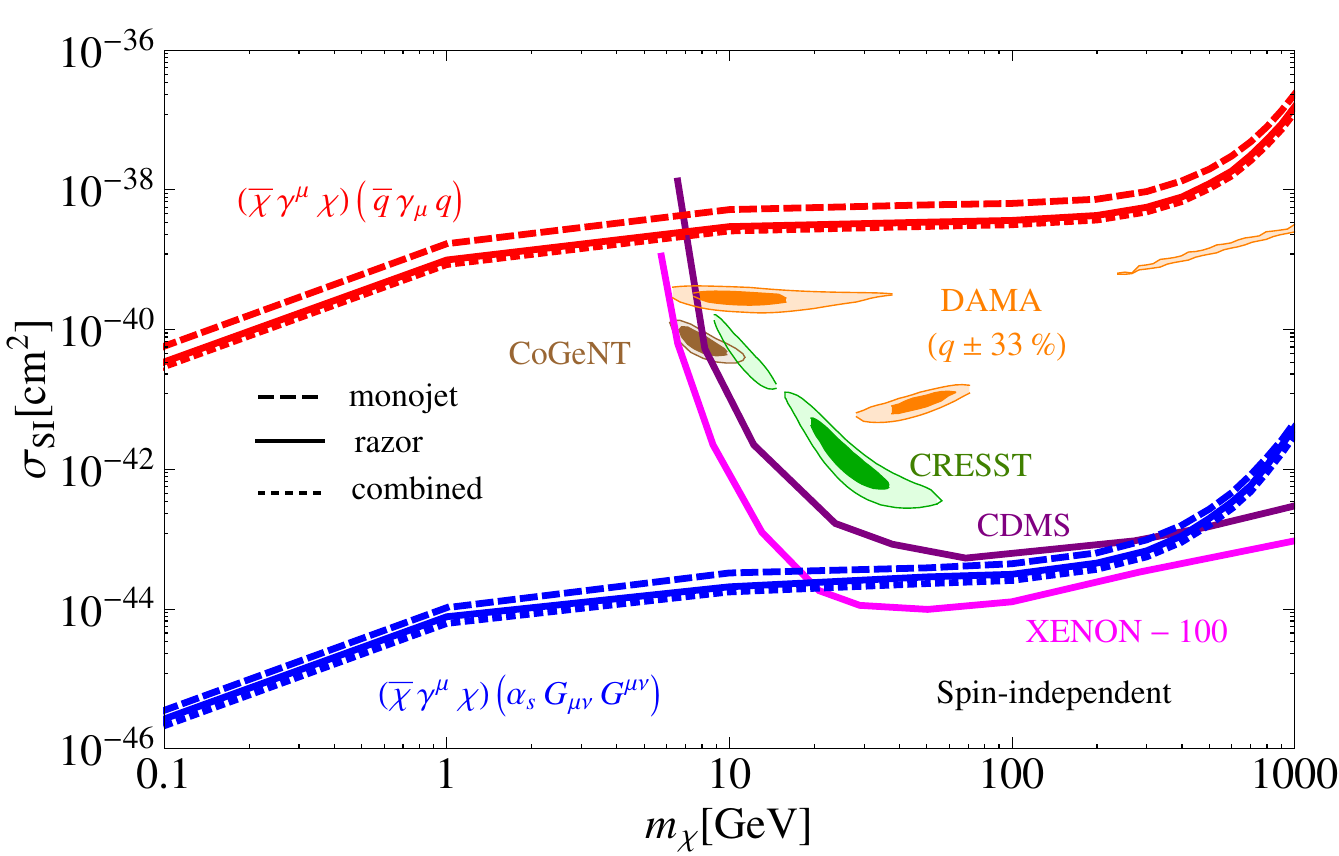}
\label{fig:spinindependent}
\includegraphics[width=0.48\textwidth]{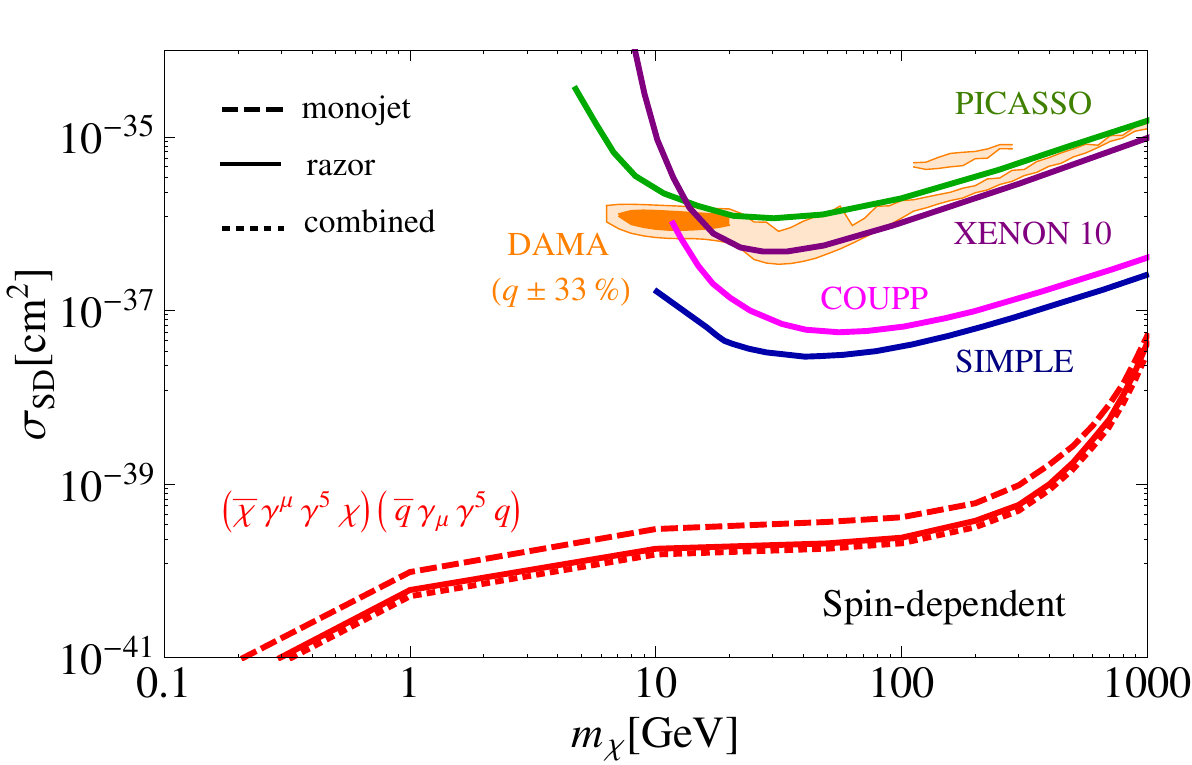}
\label{fig:spindependent}
\caption{Razor limits on spin-independent (LH plot) and spin-dependent (RH plot) DM-nucleon scattering compared to limits from the direct detection experiments. We also include the monojet limits and the combined razor/monojet limits. We show the constraints on spin-independent scattering from CDMS \cite{Ahmed:2009zw}, CoGeNT \cite{Aalseth:2010vx}, CRESST \cite{Angloher:2011uu}, DAMA \cite{Bernabei:2008yi}, and XENON-100 \cite{Aprile:2011hi}, and the constraints on spin-dependent scattering from COUPP \cite{Behnke:2010xt}, DAMA \cite{Bernabei:2008yi}, PICASSO \cite{BarnabeHeider:2005pg}, SIMPLE \cite{Girard:2011xc}, and XENON-10 \cite{Angle:2008we}. We have assumed large systematic uncertainties on the DAMA quenching factors: $q_{\textrm{Na}}=0.3\pm0.1$ for sodium and $q_I=0.09\pm0.03$ for iodine \cite{Hooper:2010uy}, which gives rise to an enlargement of the DAMA allowed regions. All limits are shown at the 90\% confidence level. For DAMA and CoGeNT, we show the 90\% and $3\sigma$ contours based on the fits of~\cite{Kopp:2009qt}, and for CRESST, we show the $1\sigma$ and $2\sigma$ contours.}
\label{fig:directdetection}
\end{figure}

We make the simplifying assumption that the effective DM-SM couplings are universal in quark flavor. However, we can account for different $u$ and $d$ couplings (\ie~$c_u\neq c_d$, where the couplings to DM are of the form $c_{u(d)}/\Lambda^2$) by rescaling the collider limits on the DM-nucleon cross-section by a factor of $(\Lambda_u^4+\Lambda_d^4)/(c_u^2\Lambda_u^4+c_d^2\Lambda_d^4)$. The bounds on the DM-nucleon cross-sections for various operators can be found in Fig.~\ref{fig:directdetection}. From the figure, we can see that collider experiments can probe DM mass regions below direct detection experiment thresholds. In the case of spin-independent scattering, the cross section bound obtained from $\mathcal O_G$ is 2-3 orders of magnitude below the cross-sections required to fit the excesses seen at DAMA, CoGeNT and CRESST. Moreover, the bound for $\mathcal O_G$ is competitive with the cross-section bounds obtained from CDMS and XENON experiments. The DM-nucleon spin-dependent scattering is not coherent over the whole nucleus, therefore the cross section bounds from spin-dependent experiments are lower then the bounds from spin-independent experiments. In this case, the collider experiments provide the strongest bound up to DM masses of $\sim$ 1 TeV. The collider bounds weaken rapidly for higher DM mass since the center-of-mass energy required to create a pair of DM is higher.

\begin{figure}[t]
\begin{center}
\includegraphics[width=0.65\textwidth]{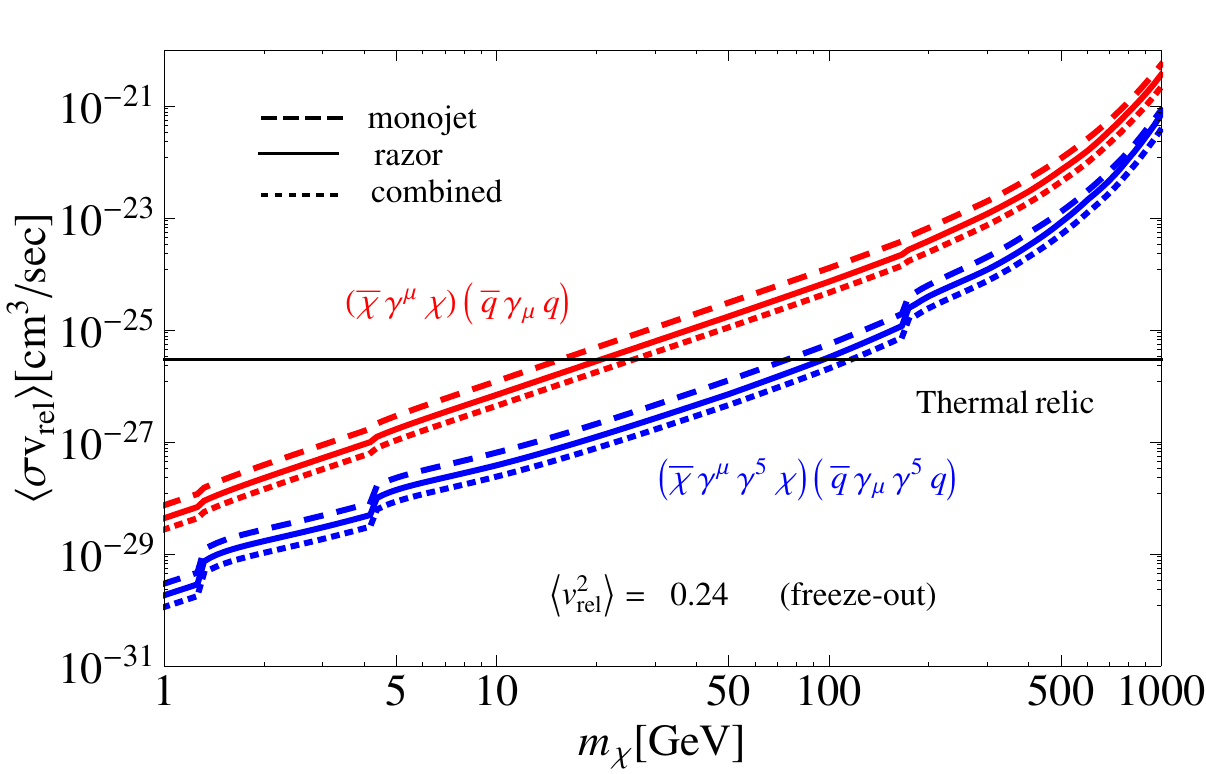}
\caption{Razor constraints on DM annihilation for flavor-universal vector or axial couplings of DM to quarks. We set $\langle v_{\textrm{rel}}^2\rangle=0.24$ which corresponds to the epoch when thermal relic DM freezes out in the early universe. However, $\langle v_{\textrm{rel}}^2\rangle$ is much smaller in present-day environments (\ie~galaxies) which results in improved collider bounds on the annihilation rate. The horizontal black line indicates the value of $\langle v_{\textrm{rel}}^2\rangle$ required for DM to be a thermal relic.}
\label{fig:annihilation}
\end{center}
\end{figure}

In addition to the direct detection bounds, we can also convert the collider bounds into a DM annihilation cross-section, which is relevant to DM relic density calculations and indirect detection experiments. The annihilation rate is proportional to the quantity $\langle \sigma v_{rel}\rangle$, where $\sigma$ is the DM annihilation cross section, $v_\mathrm{rel}$ is the relative velocity of the annihilating DM and $\langle . \rangle$ is the average over the DM velocity distribution. The quantity $\sigma v_\mathrm{rel}$ for $\mathcal O_V$ and $\mathcal O_A$ operators is \footnote{A comprehensive study of different types of operators can be found in Ref.~\cite{Goodman:2010ku}.}
\begin{eqnarray}\label{eq:annihiliationV}
\sigma_V v_{\textrm{rel}}&=&\frac{1}{16\pi\Lambda^4}\sum_q\sqrt{1-\frac{m_q^2}{m_{\chi}^2}}\left(24(2m_{\chi}^2+m_q^2)+\frac{8m_\chi^4-4m_\chi^2m_q^2+5m_q^4}{m_\chi^2-m_q^2}v_{\textrm{rel}}^2\right)\\ \label{eq:annihiliationA}
\sigma_A v_{\textrm{rel}}&=&\frac{1}{16\pi\Lambda^4}\sum_q\sqrt{1-\frac{m_q^2}{m_{\chi}^2}}\left(24m_q^2+\frac{8m_\chi^4-22m_\chi^2m_q^2+17m_q^4}{m_{\chi}^2-m_q^2}v_{\textrm{rel}}^2\right)
\end{eqnarray}
As in the case of direct detection, we assume universal DM couplings in quark flavor. In Fig.~\ref{fig:annihilation}, we show $\langle\sigma v_\textrm{rel}\rangle$ as functions of the DM mass, taking $\langle v_{rel}^2 \rangle = 0.24$, which corresponds to the average DM velocity during the freeze-out epoch. A much smaller average $\langle v_{rel}^2 \rangle$, e.g. in the galactic environment, would lead to stronger bounds. If the DM has additional annihilation modes, the bounds weaken by a factor of $1/\textrm{BR}(\bar \chi \chi \rightarrow \bar q q)$. Assuming that the effective operator description is still valid during the freeze-out epoch, the thermal relic density cross-section is ruled out at 90 \% C.L. for $m_\chi \lesssim 20 \textrm{ GeV}$ for $\mathcal O_V$, and $m_\chi \lesssim 100 \textrm{ GeV}$ for $\mathcal O_A$.

\section{Beyond Effective Theory}\label{sec:lightmediators}

So far we have made the assumption that the effective theory valid at direct detection experiments, where the typical momentum transfer is of order 100 MeV, is also valid for calculating cross sections at the LHC, where the relevant scales are of order hundreds of GeV to a TeV. Given the large hierarchy between the scales probed at the two classes of experiments it is important to consider the possibility that this assumption is violated. In particular, the presence of new particles at or below the LHC scale can modify the bounds. In fact, the disparity between these scales is so large that it has been argued that due to unitarity limits, new physics beyond the DM particle \emph{must} lie within the LHC's kinematic reach in order to generate direct detection cross sections as large as those discussed in the previous sections
~\cite{Shoemaker:2011vi}. In this section, we will investigate these issues. We shall see that even if a new mediator must be within the LHC's reach, for DM masses below a couple of hundred GeV the mediator can easily be sufficiently heavy that it does not significantly affect the search in question.
We will also find that when the new mediator is sufficiently light to modify the bounds 
the limits derived so far may be either strengthened or weakened, depending on the mass of the mediator relative to the LHC scale and relative to the mass of the DM particle. 
The issue of light mediators and how they affect mono-jet and mono-photon bounds on DM has also been discussed in~\cite{Bai:2010hh, Fox:2011fx, Fox:2011pm, Friedland:2011za, Goodman:2011jq,An:2012va}.  Furthermore, if the mediator is light it can also be searched for directly by looking for a dijet resonance or the dijet angular distribution~\cite{An:2012va}.

\subsection{Unitarity}

In~\cite{Shoemaker:2011vi}, it was shown that unitarity of $q\bar q$ forward scattering with a center of mass energy of $\sqrt{\hat s}$ places a limit on the production of DM at that energy. In particular, this argument places a lower bound on the cutoff scale $\Lambda$ 
\begin{equation}
\Lambda\gtrsim 0.4 \sqrt{\beta(\hat s) \hat s}
\label{eq:unitarity}
\end{equation}
where $\beta$ is the DM velocity which is always of order one and will hence be ignored.
In~\cite{Shoemaker:2011vi}, it was argued that an approximate requirement for the effective theory to be valid at the LHC is that this bound be satisfied at $\sqrt{\hat s}=\sqrt{s_*}$ which was set to 7 TeV.  However, this requirement is not directly related to the search in question, as both our razor analysis and the monojet searches in~\cite{Fox:2011pm, Rajaraman:2011wf}, do not probe scales of 7 TeV.

We wish to make direct contact between the unitarity limit in Eq.~\ref{eq:unitarity} and an actual collider search for DM. The first difficulty is that the unitarity argument places a limit on DM pair production at $\sqrt{\hat s}$ as opposed to DM plus any number of jets. The former does not yield observable signals at the collider. In order to make contact with more inclusive searches it is useful to interpret the limit in Eq.~\ref{eq:unitarity} as a limit, not on the energy of the incoming quarks, but on the center of mass energy of the DM system, $m_{\chi\chi}$. For the exclusive process, $q\bar q \to \chi\bar\chi$, these two scales are obviously the same, but in an inclusive process, $q\bar q \to \chi\bar\chi+X$, it is not.
This amounts to replacing the $\sqrt{\hat s}$ by the invariant mass of the DM system $m_{\chi\chi}$, or
\begin{equation}
\label{eq:limit-mchichi}
m_{\chi\chi}<\frac{\Lambda}{0.4}\,.
\end{equation}
This substitution allows us to make contact with any DM production process being probed at the collider. 

\begin{figure}[t]
\begin{center}
\subfigure[~$m_\chi=1$ GeV]{
\includegraphics[width=0.48\textwidth]{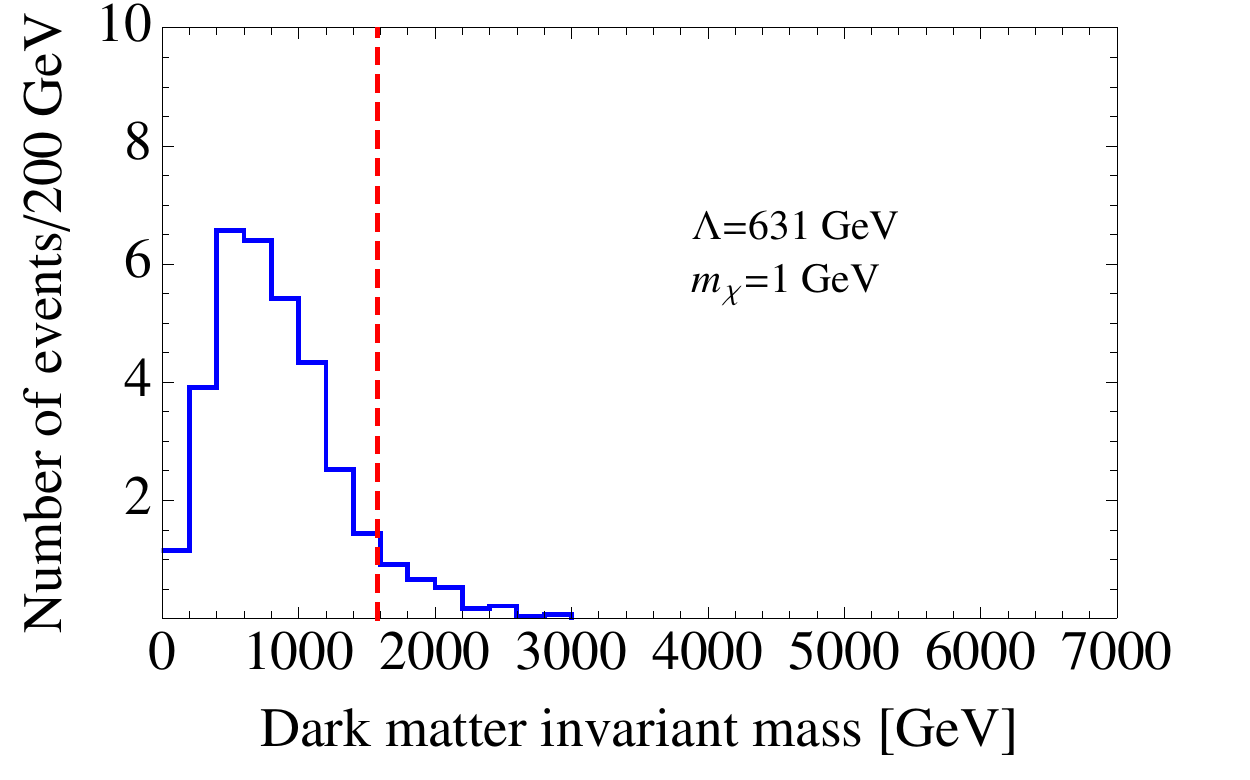}}
\subfigure[~$m_\chi=100$ GeV]{
\includegraphics[width=0.48\textwidth]{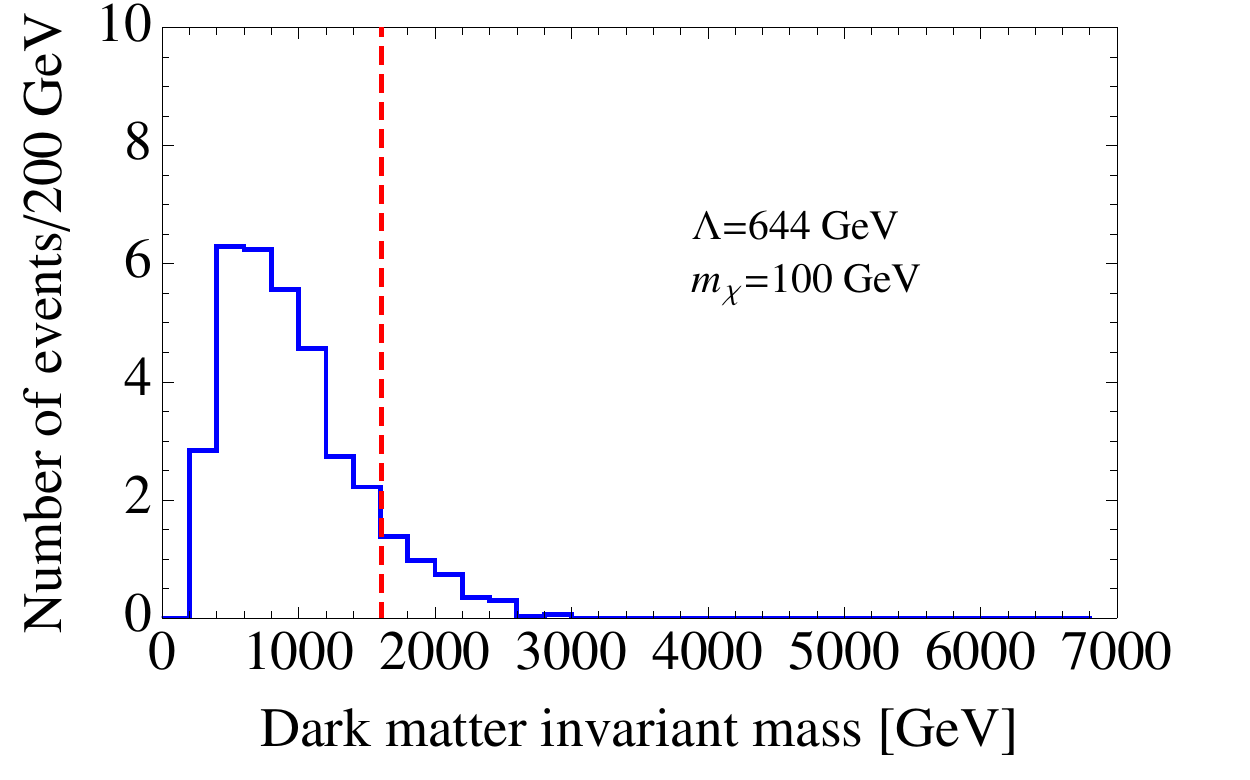}}
\subfigure[~$m_\chi=500$ GeV]{
\includegraphics[width=0.48\textwidth]{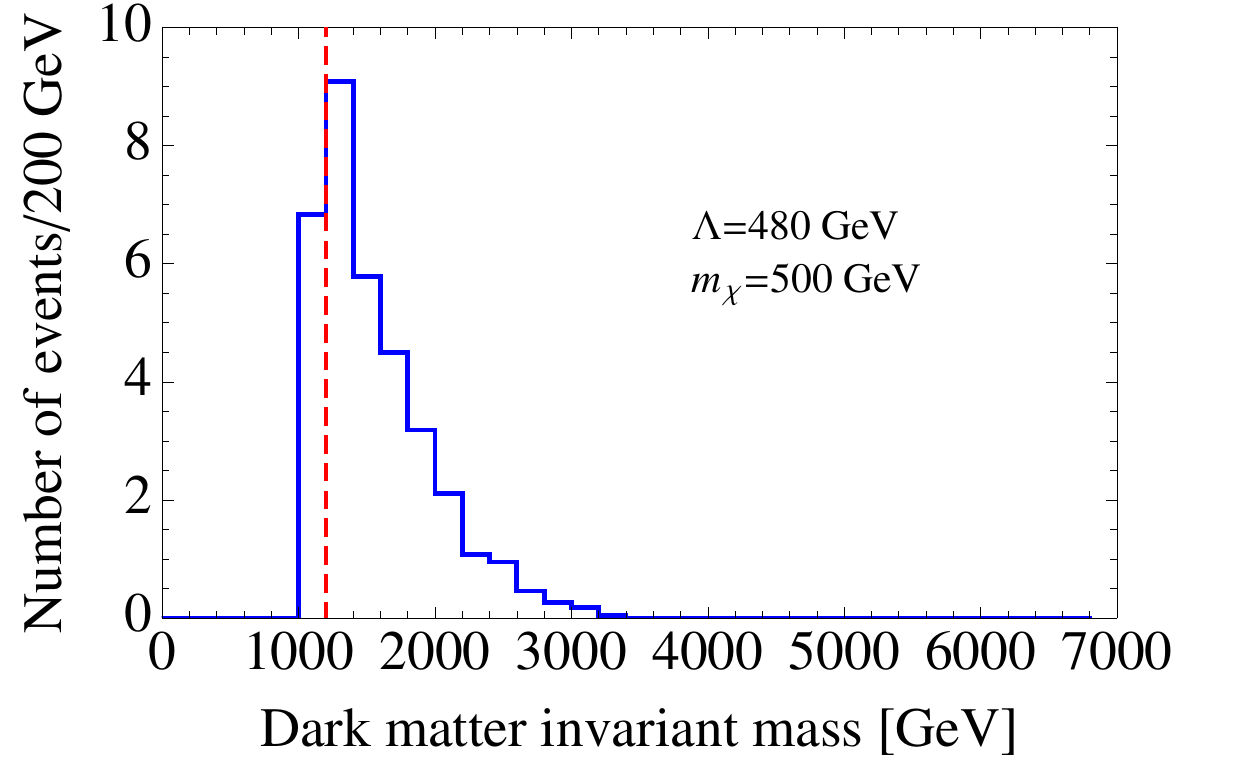}}
\caption{$m_{\chi\chi}$ distribution for signal events with $u$-quark vector couplings with $R^2>0.81$ and $M_R>250$ GeV. The red dashed line corresponds to the unitarity bound $m_{\chi\chi}=\Lambda/0.4$. The three panels show the distribution for DM masses of (a) 1~GeV, (b) 100~GeV, and (c) 500 GeV.  The fractions of events which lie beyond the bound are 8\%, 11\% and 80\% respectively.}
\label{fig:mchichi}
\end{center}
\end{figure}

We can now ask the following question. Assuming a contact interaction of quarks with DM with a cutoff scale $\Lambda$ right at where we have set our limits, what fraction of the signal events violate Eq.~\ref{eq:limit-mchichi} ? In Fig.~\ref{fig:mchichi} we show the invariant mass distribution of events passing our analysis cuts for a few DM masses. 
We show the unitarity limit of $\Lambda/0.4$ as a dashed vertical line. Events that violate the bound are guaranteed to be sensitive to the physics that mediates the interaction of quarks and DM, and thus are not reliably described by the effective theory. Events that are to the left of the vertical line may be described by the effective theory, (unless the mediator is light, see below).
For DM masses of 1 and 100 GeV, the fraction of events that violate the unitarity limit is 8\% and 11\% respectively. Thus, the bound derived with the full effective theory may be accurate to within this precision, which we consider acceptable. The situation is different for heavier DM, \eg~500 GeV. Here, the fraction of ``unitarity violating'' events is high at 80\%. This is due to two effects. First, the scale $\Lambda$ which the analysis constrains (see Figs.~\ref{fig:DM_bounds} and~\ref{fig:combined}), and hence the unitarity limit, is lower. In addition, the invariant mass distribution is pushed to higher values of $m_{\chi\chi}$ due to the higher threshold. 

We thus conclude that the effective theory can be valid for DM masses below a few hundred GeV, where the limit on $\Lambda$ is still flat. This conclusion is in qualitative agreement with previous analyses~\cite{Beltran:2010ww, Fox:2011pm} which used arguments of perturbativity rather than unitarity. We emphasize that, as we shall see in the next subsection, the cross section can deviate from that derived via effective theory if the mediator is light, within the reach of the analysis. As the mass of the DM becomes heavy enough so that its production is kinematically suppressed by parton distribution functions (PDFs), the effective theory description breaks down and the UV physics must be accounted for in order to get an accurate description of the limits. In the next subsection we will consider a simplified model which includes the mediating particles explicitly and investigate how the bounds are modified. We will also see that requiring perturbative simplified models gives qualitatively similar results to the requirements of unitarity.

\subsection{Light Mediators}

We now replace the effective theory analyzed above for a renormalizable ``simplified" model. Consider a neutral vector particle of mass $M$ which couples to DM pairs with a coupling of $g_\chi$ and to up-quarks with a coupling of $g_q$. At low energies, say those relevant for direct detection, this model is described well by an effective theory with a vector operator suppressed by the scale $\Lambda\equiv M/\sqrt{g_\chi g_q}$.

If the mediator is sufficiently light, but still heavier than $2m_\chi$ the mediator may be produced on-shell, and subsequently decay to a pair of DM particles.
This leads to an enhanced production rate proportional to $g_q^2 g_\chi^2 / (M \Gamma)$ where $\Gamma$ is the total width of the mediator particle.
If the mediator is much lighter than twice the DM mass, the DM production is proportional to $g_q^2 g_\chi^2/m_{\bar{\chi}\chi}$ and is significantly suppressed.

\begin{figure}[t]
\begin{center}
\includegraphics[width=0.48\textwidth]{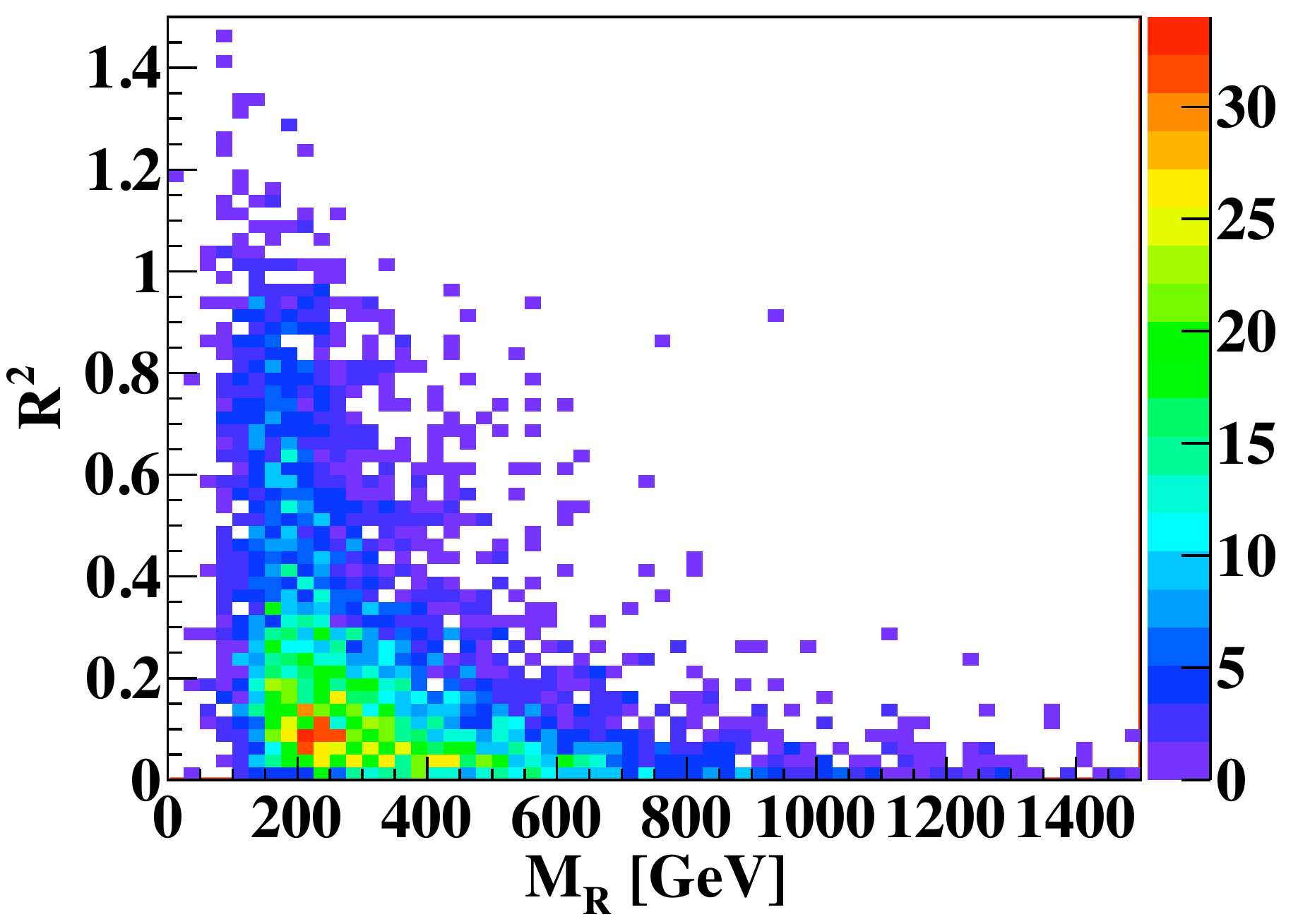}
\includegraphics[width=0.48\textwidth]{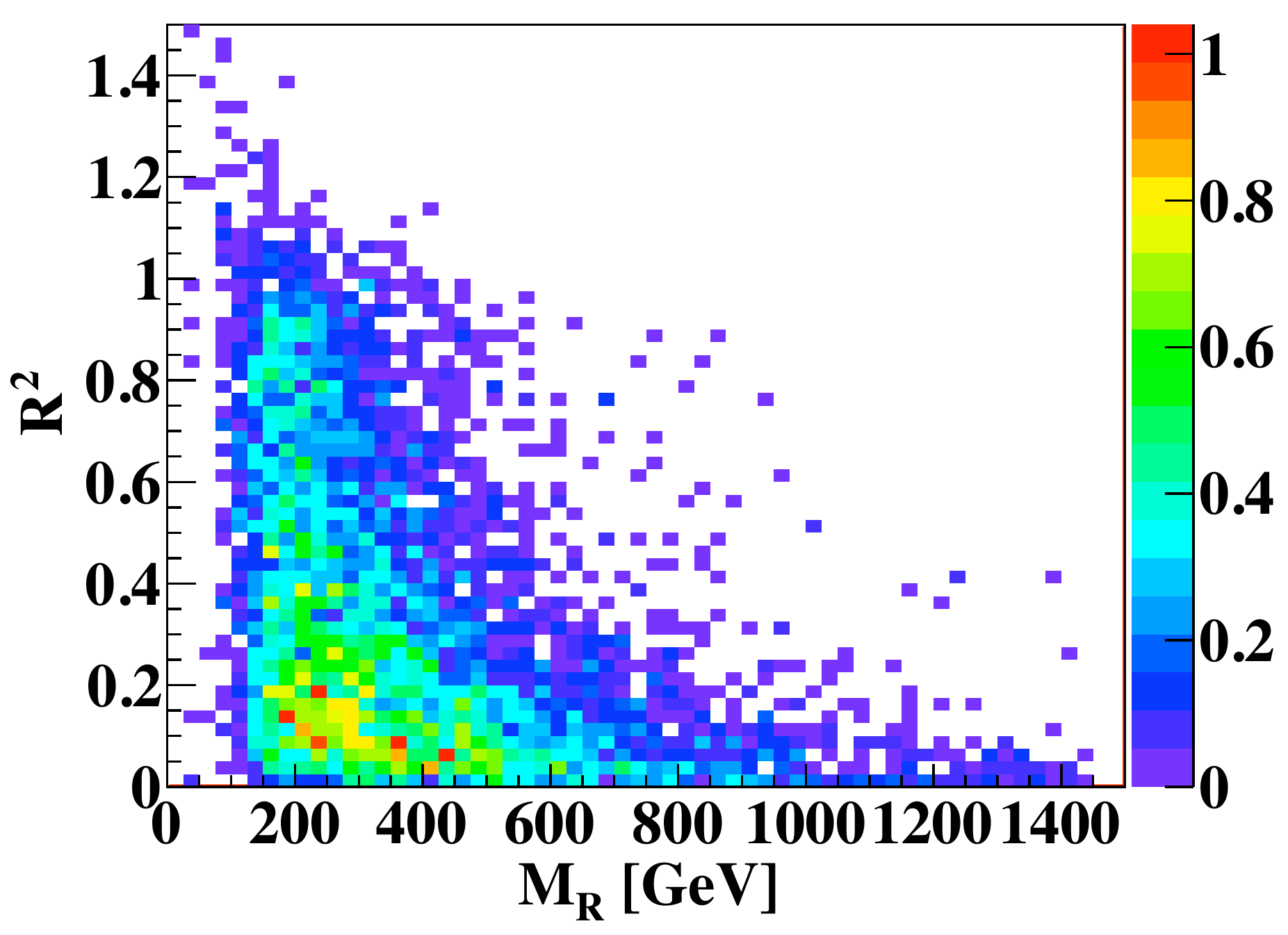}
\caption{$R^2$ vs. $M_R$ for light mediators, with arbitrary normalization.  The LH plot corresponds to the case of $m_{\chi}=50$ GeV, $M_{Z'}=100$ GeV, $\Gamma_{Z'}=M_{Z'}/3$ and the RH plot to $m_{\chi}=50$ GeV, $M_{Z'}=300$ GeV, $\Gamma_{Z'}=M_{Z'}/3$.}
\label{fig:rmr_lightmediators}
\end{center}
\end{figure}

The presence of a light mediator can also affect the kinematic distribution of the signal. In particular, in the case of on-shell production of a mediator which decays to DM, one would expect the signal to be quite similar to the background of on-shell production of a $Z$ which decays invisibly. Indeed, in Fig.~\ref{fig:rmr_lightmediators} we show the distribution of $M_R$ and $R^2$ for a mediator masses of 100~GeV and 300~GeV, and a DM mass of 50~GeV. One can see that the congregation of events around $R^2\sim 1$ is absent and the distribution is similar to that of the $Z+$ jets background (see Fig.~\ref{fig:rmr_zjets}). As a result, the cut efficiency for this case will be lower, which will partially counter the gain in overall rate when calculating the ultimate bounds.

\begin{figure}[t]
\begin{center}
\includegraphics[width=0.65\textwidth]{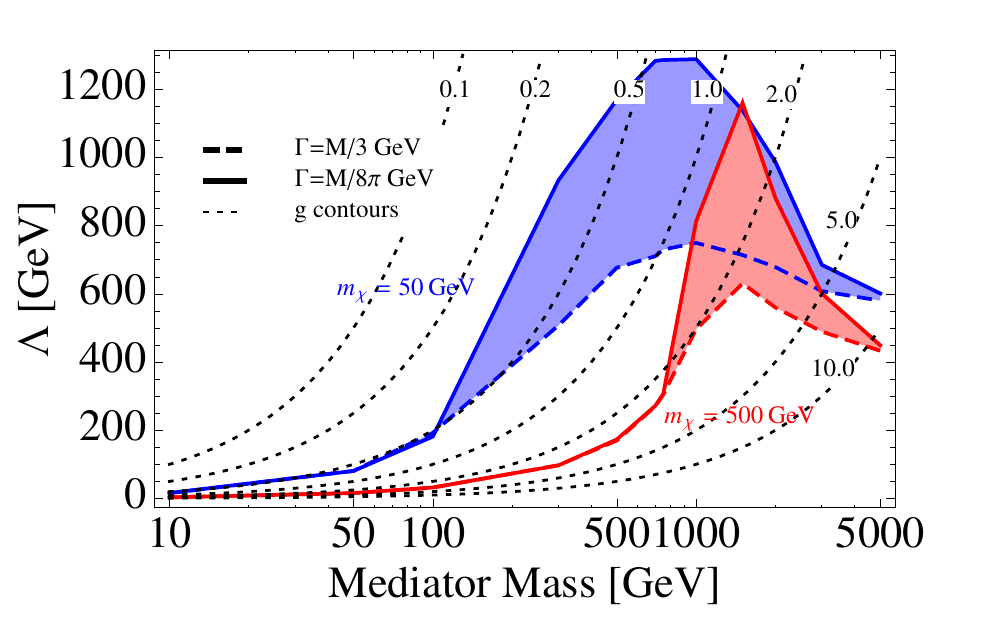}
\caption{Cutoff scale $\Lambda\equiv M/g$ bounds as a function of mediator mass $M$, where $g\equiv \sqrt{g_\chi g_q}$. We assume $s$-channel vector-type interactions and consider DM masses of $m_\chi=50$ GeV (blue) and $m_\chi=500$ GeV (red). We vary the width $\Gamma$ of the mediator between $M/3$ (solid line) and $M/8\pi$ (dashed line).
\label{fig:lightmediators}}
\end{center}
\end{figure}

In Fig.~\ref{fig:lightmediators}, we show the limits we achieve on $\Lambda\equiv M/\sqrt{g_q g_\chi}$ as a function of the mediator mass $M$ for two fixed DM masses, 50 and 500 GeV. For each case, we consider a range of widths for the mediator between $M/3$ and $M/8\pi$. We consider these two values as extremes of what is possible in general, although the narrow width may not be physically realizable for the DM couplings we consider here.  We see that as the mediator mass is lowered the bound improves because DM production proceeds through the production of an on-shell mediator which later decays. The improvement can be substantial, as much as a factor of 5 in the limit on the cross section in the narrow mediator case. As the mediator mass is lowered further and its mass drops below threshold for DM production the limit weakens significantly, as expected.

We conclude that while it is easy for physics beyond the DM effective theory to modify the bounds derived within the effective theory, this modification can either cause bounds to improve in the intermediate mediator mass region or to weaken in the light mediator region. 
\section{Discussion and Future Prospects}\label{sec:conclusions}

In this paper, we expand on previous work done on DM limits at colliders using monojets by utilizing the razor analysis of CMS. At the LHC, one expects events that contain several high $p_T$ jets, and the monojet requirement may restrict the signal efficiency. By allowing for an arbitrary number of hard jets, we can improve upon the signal efficiency. Furthermore, the razor analysis uses a complementary data set to that of the monojet search, thus allowing one to combine the bounds from the two methods. 

We estimate the razor bounds on dark matter that one could expect to achieve after approximately $800 \text{ pb}^{-1}$ of LHC data, and find that they are slightly better than those from the existing monojet search, which is based on $1 \text{ fb}^{-1}$.  The improvement is about 40\% in the direct detection cross section.
Since the uncertainties of the razor analysis are dominantly statistical in nature we expect this bound to improve with further updates of the razor analysis employing larger data sets.  
 
We also address the validity of using an effective theory. We find that for light DM masses (below a few hundred GeV), the bound derived using an effective theory may be accurate to about 10\%. If the mediator is heavy, but below a couple TeV, the limit derived from effective theory is in fact conservative, and the true limit is stronger. But, if the mediator is too light to decay to dark matter pairs the true limit is far weaker than the one derived from effective theory.
In addition, we find that the effective theory breaks down at DM masses that are heavy enough such that the DM production is kinematically suppressed by PDFs, and we must take into account the UV physics in these cases. 
 
Although originally conceived of as a search tool for squarks/gluinos in supersymmetry we have demonstrated that razor analysis is a powerful technique to also look for production of non-colored states that lead to missing energy in the detector.  The ease with which it discriminates between signal and background makes us optimistic for future, dedicated analyses, to search for DM that use this technique.  Furthermore, should an excess be observed, the existence of additional observables beyond those available in monojet/monophoton searches may prove beneficial in its interpretation.

\section*{Acknowledgements}
The authors wish to thank John Campbell, Marat Freytsis, Joe Lykken, Adam Martin, Yuhsin Tsai, and Felix Yu for many helpful discussions and Joachim Kopp for insightful comments on a early version of this manuscript.  We thank Ciaran Williams for his patience in explaining how to use MCFM. RP and C.-T.Y are supported by the Fermilab Fellowship in Theoretical Physics. RP thanks the National Science Foundation for support under Grants PHY-0757481 and PHY-1068008. Fermilab is operated by Fermi Research Alliance, LLC, under Contract DE-AC02-07CH11359 with the United States Department of Energy.

\bibliographystyle{apsrev}
\bibliography{razorDM_v3}

\end{document}